\documentclass[12pt]{elsarticle}
\usepackage{amsfonts,enumerate,amsmath,amssymb}
\usepackage{color,latexsym,amsfonts,amssymb}
\usepackage{ulem}
\usepackage{appendix}
\usepackage{ragged2e}

\newcommand{\bed}{\begin{displaymath}}
\newcommand{\eed}{\end{displaymath}}
\newcommand{\bea}{\bed\begin{array}{rl}}
	\newcommand{\eea}{\end{array}\eed}

\newcommand{\barray}{\begin{array}{ll}}
	\newcommand{\earray}{\end{array}}
\numberwithin{equation}{section}

\begin{document}
\begin{frontmatter}	
\title{Rate-dependent tipping-delay phenomenon in a thermoacoustic system with colored noise}

\author[author1]{Xiaoyu Zhang}
\ead{xiao\_yu\_zhang@yahoo.com}

\author[author1,author2]{Yong Xu\corref{mycorrespondingauthor}}
\cortext[mycorrespondingauthor]{Corresponding author}
\ead{hsux3@nwpu.edu.cn}

\author[author1]{Qi Liu}
\ead{liuqi1780280327@yahoo.com}

\author[author3]{J{\"u}rgen Kurths}
\ead{Juergen.Kurths@pik-potsdam.de}

\address[author1]{Department of Applied Mathematics, Northwestern Polytechnical University, Xi'an, 710072, China}

\address[author2]{MIIT Key Laboratory of Dynamics and Control of Complex Systems, Northwestern Polytechnical University, Xi’an, 710072, China}

\address[author3]{Potsdam Institute for Climate Impact Research, Potsdam 14412, Germany}

\setlength{\baselineskip}{0.26in}	
\begin{abstract}
Tipping is a phenomenon in multistable systems where small changes in inputs cause huge changes in outputs. When the parameter varies within a certain time scale, the rate will affect the tipping behaviors. These behaviors are undesirable in thermoacoustic systems, which are widely used in aviation, power generation and other industries. Thus, this paper aims at considering the tipping behaviors of the thermoacoustic system with the time-varying parameters and the combined excitations of additive and multiplicative colored noises. Transient dynamical behaviors for the proposed thermoacoustic model are implemented through the reduced Fokker-Planck-Kolmogorov equation derived by a standard stochastic averaging method. Then, the tipping problems of the rate-dependent thermoacoustic systems with random fluctuations are studied by virtue of the obtained probability density functions. Our results show that the rate delays the value of the tipping parameter compared to the one with the quasi-steady assumption, which is called as a rate-dependent tipping-delay phenomenon. Besides, the influences of the initial values, the rate, the changing time of the parameters, and the correlation time of the noises on the rate-dependent tipping-delay phenomenon are analyzed in detail. These results are of great significance for research in related fields such as aviation and land gas turbines.
\end{abstract}
\end{frontmatter}

\setlength{\baselineskip}{0.26in}
\section{Introduction}

\label{sec1}
Many systems in nature have the property of multistability, that is, multiple stable states of a system coexist. In multistable systems, small changes in inputs may cause sudden and disproportionate changes in the outputs. This phenomenon is named as tipping \cite{ashwin2017parameter}, which exists widely in climatology \cite{holland2006future,zickfeld2005indian}, ecology \cite{clark2013light,hoegh2007coral,mumby2007thresholds} and economics \cite{yan2010diagnosis}, etc. According to the inherent mechanism of tipping, Ashwin et al. classified it into three categories: bifurcation-induced tipping (B-tipping), noise-induced tipping (N-tipping) and rate-induced tipping (R-tipping) \cite{ashwin2012tipping}. B-tipping refers to a tipping phenomenon that occurs when the value of external force or internal parameters slowly exceeds a critical threshold, resulting in a bifurcation behavior. N-tipping is a phenomenon where the system switches its state after being disturbed by noise \cite{ma2018detecting, ma2019predicting}. R-tipping is a new concept proposed in recent years. For such a case, the parameters of the system are no longer static relative to the variables, but become a new time-dependent variable. When the rate of parameters exceeds a critical value, R-tipping occurs. In other words, R-tipping describes the phenomenon that the parameters change too fast and the system can not adapt to it, which causes the system undergoing a qualitative change. Although R-tipping has frequently entered the field of vision in recent years, the phenomena in nature are abound \cite{lucarini2005destabilization, mitry2013excitable}. Among them, the most concerned is the global warming \cite {lenton2019climate}, which focuses on the rate of warming rather than the temperature itself.

In fact, it is highly desirable to consider the rate of parameters in a system. In traditional approaches of dynamics research, some variables with relatively insignificant changes are often regarded as control parameters with some fixed value for simplicity. However, in nature, the dynamical systems often strongly depend on time, and the parameters will change in a certain time scale, which will affect the dynamic states of the systems. With the deepening of the follow-up research, influences of the time-varying parameters must be considered in order to describe the more accurate dynamical behaviors. When the rate is introduced into the system, the values of parameters caused tipping will be delayed compared with that of a quasi-steady case, that is, the particle will hover around the previous state because of inertia. We called this effect as ‘rate-dependent tipping-delay phenomenon’.

A representative example considering the rate of parameters is the thermoacoustic system, which is closely related to the burner devices in rocket and aeroengine. In such systems, tipping occurs when there is a positive feedback between the fluctuation of sound pressure and unsteady heat release rate in the combustion chamber, which is named as thermoacoustic instability \cite{lieuwen2012unsteady}. Thermoacoustic instability causes a pressure oscillation with a large amplitude in the combustion chamber, which may lead to a sudden drop in aircraft altitude and a loss of productivity of onshore gas turbines. In the study of Apollo in the 1960s, billions of dollars were spent to find solutions to mitigate thermoacoustic instability \cite{oefelein1993comprehensive}. Therefore, understanding  thermoacoustic instability is still a challenging and urgent problem. It was found that tipping between thermoacoustic instability and other states is a time-varying process with intermittent properties \cite{nair2014multifractality}, and fluctuations of the equivalence ratio can cause the thermoacoustic instability \cite{lieuwen1998role}. Thus, the equivalent ratio is considered as a time-varying parameter to investigate the transition behaviors of the thermoacoustic systems between different system states and the rate-dependent tipping-delay phenomenon in this study. Because most of the thermoacoustic systems need to be loaded or unloaded quickly, the tipping-delay phenomenon caused by the rate could bring the system  into danger \cite{bonciolini2018experiments}. For this reason, a in-depth exploring of the rate-dependent tipping-delay phenomenon in  thermoacoustic systems is uttermost important in aviation, land gas turbine and other related fields.

Random disturbances are ubiquitous and play an important role in the evolution of dynamical systems in the macro world. Throughout the various strands of research, the influences of random disturbances on dynamical systems have become a frontier topic and developed a wide-ranging corpus of results \cite{zhang2019random,mei2019transport,wang2016levy,liu2020bistability}. Any systems are inevitably disturbed by uncertainties, and thermoacoustic systems are no exception. The fluctuation of the heat release rate caused by incoherent turbulence in thermoacoustic systems is often represented by noise \cite{bonciolini2018experiments}. As is well-known, white noise is the most commonly used noise in nonlinear dynamics. Recently, Bonciolini et al. studied the dynamical behaviors of a thermoacoustic model under additive white noise \cite{bonciolini2018experiments}. They verified that the tipping-delay phenomenon occurs in rate affecting systems by experimental observations and numerical results. In addition, Unni et al. discussed the effect of initial conditions of  thermoacoustic systems driven by additive white noise on the tipping-delay phenomenon \cite{unni2019interplay}. The aforementioned works mainly focused on dynamical systems with additive Gaussian white noise excitation. But in fact, the actual fluctuation always has a non-zero correlation time and a non-constant spectral distribution, which is discribed by colored noise. This suggests that it is necessary to take colored noise into account in the study of nonlinear systems \cite{li2019transition,liu2018sliding}. Apart from the additive case, the multiplicative noise which is related to the state of the system also plays an important role in this research. Thus, this paper explores the tipping-delay phenomenon in  thermoacoustic systems disturbed by additive and multiplicative exponential colored noises, which can be regarded as a substantial generalization of the previous works.

The aim of this paper is to study the transient dynamical behavior of a class of thermoacoustic systems with time varying parameters excited by additive and multiplicative colored noise fluctuations. Based on the standard stochastic averaging method, we consider the rate-dependent tipping-delay phenomenon in the cases of additive colored noise, multiplicative colored noise, and combined additive and multiplicative colored noises respectively. Moreover, the effects of initial values, ramp rate, changing time of parameters, and the noise parameters on the transient dynamical behaviors are examined throungh the transient probability density functions (PDFs). Meanwhile, several differences between the case of colored noise and white noise are discussed.

This paper is organized as follows. In Section 2, a theoretical model of a thermoacoustic system is introduced, which is modeled as a single-degree-of-freedom Duffing-Van der Pol system with fluctuations. Besides, its transient dynamics are obtained by virtue of a standard stochastic averaging technique. Then, the influences of several parameters on the tipping-delay phenomenon are examined in Section 3. Section 4 discusses the dependence of the correlation time of noises on the dynamical behaviors. Finally, several conclusions are given to close this paper in Section 5.

\section{A thermoacoustic model with colored noise}
\label{sec2}
Thermoacoustic behaviors can be represented by the following Helmholtz equation \cite{bonciolini2018experiments, lieuwen2012unsteady, noiray2017method}
\begin{equation}\label{H eq}
\frac{\partial^{2} x}{\partial t^{2}}-c^{2} \nabla^{2} x=(\gamma-1) \frac{\partial q}{\partial t},
\end{equation}
where $x$, $q$, $c$, and $\gamma$ denote the acoustic pressure, the fluctuating component of the heat release rate $\hat{Q}=\overline{Q}+q$, the speed of sound and the heat capacity ratio, respectively. Equation (\ref{H eq}) can be characterized into the Laplace-transformed acoustic pressure field $\hat{x}$ in the combustor $V$ with the boundary condition on the surface $\sigma$
\begin{equation*}
	\begin{aligned}
		\nabla^{2} \hat{x}(s, y)-\left(\frac{s}{c}\right)^{2} \hat{x}(s, y)&=-s \frac{(\gamma-1)}{c^{2}} \hat{Q}(s, y) ,\\
		\frac{\hat{x}(s, y)}{\hat{\mathbf{u}}(s, y) \cdot \mathbf{n}}&=Z(s, y).
	\end{aligned}
\end{equation*}
Here, $s$ denotes the Laplace variable, $y$ is the position, $\hat{\mathbf{u}}$ is the acoustic velocity, $\mathbf{n}$ is the unit vector normal to the boundary and $Z$ is the impedance.
In a thermoacoustic system, the dominant mode can approximate the dynamics of the original system. Thus, we  project the acoustic field on an orthogonal basis $\psi$, with the approximation $\hat{x}(s, y) \approx \hat{\kappa}(s) \psi(y)$, yielding
\begin{equation}\label{basis}
\hat{\kappa}=\frac{s \rho c^{2}}{s^{2}+\omega_{0}^{2}} \frac{1}{V \Lambda}\left(\frac{\gamma-1}{\rho c^{2}} \int_{V} \hat{Q}(s, y) \psi(y) d V-\hat{\kappa} \int_{\sigma} \frac{|\psi(y)|^{2}}{Z(s, y)} d \sigma\right),
\end{equation}
where $\rho$ is the gas density and $\Lambda$ denotes the mode normalization coefficient. Equation (\ref{basis}) is then equivalent to the following form
\begin{equation*}
	\left(s^{2}+\alpha s+\omega_{0}^{2}\right) \hat{\kappa}=s \hat{q},
\end{equation*}
with the resonance frequency $\omega_0$, $\alpha={\rho c^{2}}/{V \Lambda} \int_{\sigma} {|\psi(y)|^{2}}/{Z(s, y)} d\sigma $ and $\hat{q}={(\gamma-1)}/{V \Lambda} \int_{V} \hat{Q}(s,y)\\
 \psi(y) d V $. 
Based on \cite{lieuwen2012unsteady,zinn1971application}, we consider the effects of the nonlinearity $\hat{q}$ up to the fifth order and cubic stiffness nonlinearity, which leads to the following thermoacoustic oscillation model with random fluctuations

\begin{equation} \label{main eq}
\ddot{x}-\left(v+\beta_{1} x^{2}-\beta_{2} x^{4}\right) \dot{x}+\omega_{0}^{2} x+\beta_{0} x^{3}=\eta(t)+x \xi(t),
\end{equation}
where $v$ is related to the linear growth rate, $\beta_0$, $\beta_1$, $\beta_2$ are real positive parameters.The nonlinear stiffness coefficient $\beta_{0}$ is assumed to be a small quantity describing the anisochronicity of the oscillations. The terms $\eta(t)$ and $\xi(t)$ are mutual independent Gaussian colored noises with zero mean and the following correlation functions
\begin{equation*}
	\begin{aligned}
		{\langle\eta(t)\rangle}&=\langle\xi(t)\rangle= 0, \\ {\langle\eta(t) \eta(s)\rangle}&=\frac{D_{1}}{\tau_{1}} \exp \left(-\frac{|t_1-t_2|}{\tau_{1}}\right), \\ {\langle\xi(t) \xi(s)\rangle}&=\frac{D_{2}}{\tau_{2}} \exp \left(-\frac{|t_1-t_2|}{\tau_{2}}\right),
	\end{aligned}
\end{equation*}
in which $D_i$ and $\tau_i$ $(i=1,2)$ are the noise intensity and correlation time of the additive noise $\eta(t)$ and the multiplicative noise $\xi(t)$ respectively. It satisfies $D_1, D_2\geq0$ but does not include that $D_1$ and $D_2$ are both zero. System (\ref{main eq}) is a subcritical Hopf bifurcation model, and its steady-state dynamic behaviors and stochastic bifurcations in a general form have been discussed in detail \cite{xu2011stochastic}.

We introduce the transformation
\begin{equation*}
	\begin{aligned}
		x(t) &=A(t) \cos \theta, \\ \dot{x}(t) &=-A(t) \omega_{0} \sin \theta, 
	\end{aligned}
\end{equation*}
where $\theta=\omega_{0} t+\varphi(t)$, $A(t)$ and $\varphi(t)$ are amplitude and phase process respectively. 

Applying the stochastic averaging method, the stochastic differential equations of $A(t)$ and $\varphi(t)$ are obtained as follows

\begin{subequations}
	\begin{align}
		d A=&\left[\frac{v}{2} A+\frac{\beta_{1}}{8} A^{3}-\frac{\beta_{2}}{16} A^{5}+\frac{D_{1}}{2 \omega_{0}^{2} A\left(1+\omega_{0}^{2} \tau_{1}^{2}\right)}+\frac{3 D_{2} A}{8 \omega_{0}^{2}\left(1+4 \omega_{0}^{2} \tau_{2}^{2}\right)}\right] d t \notag\\ &+\sqrt{\frac{D_{1}}{\omega_{0}^{2}\left(1+\omega_{0}^{2} \tau_{1}^{2}\right)}+\frac{D_{2} A^{2}}{4 \omega_{0}^{2}\left(1+4 \omega_{0}^{2} \tau_{2}^{2}\right)}} d W_{1}(t), \label{SDE_A}\\ 
		d \varphi=&\left[\frac{3 \beta_{0}}{8 \omega_{0}} A^{2}-\frac{D_{2} \omega_{0} \tau_{2}}{2 \omega_{0}^{2}\left(1+4 \omega_{0}^{2} \tau_{2}^{2}\right)}\right] d t \notag\\ &+\sqrt{\frac{D_{1}}{A^{2} \omega_{0}^{2}\left(1+\omega_{0}^{2} \tau_{1}^{2}\right)}+\frac{3 D_{2}}{4 \omega_{0}^{2}\left(1+4 \omega_{0}^{2} \tau_{2}^{2}\right)}+\frac{2 D_{2} \omega_{0}^{2} \tau_{2}^{2}}{\omega_{0}^{2}\left(1+4 \omega_{0}^{2} \tau_{2}^{2}\right)}} d W_{2}(t),
	\end{align}
\end{subequations} 
where $W_1(t)$ and $W_2(t)$ are independent standard Wiener processes. We uncover that equation (\ref{SDE_A}) is independent of $\varphi$, and then one get the following Fokker-Planck-Kolmogorov (FPK) equation where the PDF $P(A,t)$ of the amplitude $A$ satisfies
\begin{equation}\label{FPK}
\frac{\partial P(A,t)}{\partial t}=-\frac{\partial}{\partial A}\left[m(A) P(A,t)\right]+\frac{1}{2} \cdot \frac{\partial^{2}}{\partial A^{2}}\left[b(A) P(A,t)\right], 
\end{equation}
with the drift and diffusion coefficients 
\begin{equation*}
	\begin{aligned}
		m(A)&=\frac{v}{2} A+\frac{\beta_{1}}{8} A^{3}-\frac{\beta_{2}}{16} A^{5}+\frac{D_{1}}{2 \omega_{0}^{2} A\left(1+\omega_{0}^{2} \tau_{1}^{2}\right)}+\frac{3 D_{2} A}{8 \omega_{0}^{2}\left(1+4 \omega_{0}^{2} \tau_{2}^{2}\right)},\\
		b(A)&=\frac{D_{1}}{\omega_{0}^{2}\left(1+\omega_{0}^{2} \tau_{1}^{2}\right)}+\frac{D_{2} A^{2}}{4 \omega_{0}^{2}\left(1+4 \omega_{0}^{2} \tau_{2}^{2}\right)}.
	\end{aligned}
\end{equation*}

The qualitative behaviors of the stochastic system (\ref{main eq}) depend on the drift and diffusion coefficients on two boundaries. The existence of steady-state PDFs (SPDFs) $P_s(A)$ can be proved by the judgment criterion in \cite{zhu2017introduction}. We find that the SPDF always exists in the cases of only additive colored noise excitation, and additive and multiplicative colored noise co-excitation. However, when only multiplicative colored noise is considered, the SPDF exists if and only if $v>-D_{2}/\left[2 \omega_{0}^{2}\left(1+4 \omega_{0}^{2} \tau_{2}^{2}\right)\right]$. Let $\partial P(A,t)/\partial t=0$, then the SPDFs can be deduced to 
\begin{equation*}
	P_s(A)=\frac{C}{b^{2}(A)} \exp \left[2 \int \frac{m(A)}{b^{2}(A)} d A\right],
\end{equation*}
where $C$ is a normalization constant which satisfies the normalization condition $\int P_s(A)dA=1$. 

\section{Rate-dependent tipping-delay phenomenon}
\label{sec3}
In this section, we study the transient dynamical behaviors of the thermoacoustic system (\ref{main eq}) especially for the rate-dependent tipping-delay phenomenon. The parameters of industrial thermoacoustic systems often change with time. To close the actual situations, the parameter $v$ is assumed to be a piece-wise linear time-varying one involving a ramp $v(t)=v_0+Rt$, where $v_0$ and $R$ denote the initial value and the ramp rate, respectively. This form of parameter variation is commonly used to characterize diverse dynamical phenomena in climate, physics and industrial systems \cite{kaszas2019tipping}. Three cases with additive colored noises, multiplicative colored noises, and combined additive and multiplicative colored noises are considered in detail separately. Besides, we also investigate systematically effects of the initial value, ramp rate and changing time of the parameters on the tipping-delay phenomenon.

\subsection{The case of additive colored noise}

\begin{figure}[h]
	\centering
	\includegraphics[width=9cm]{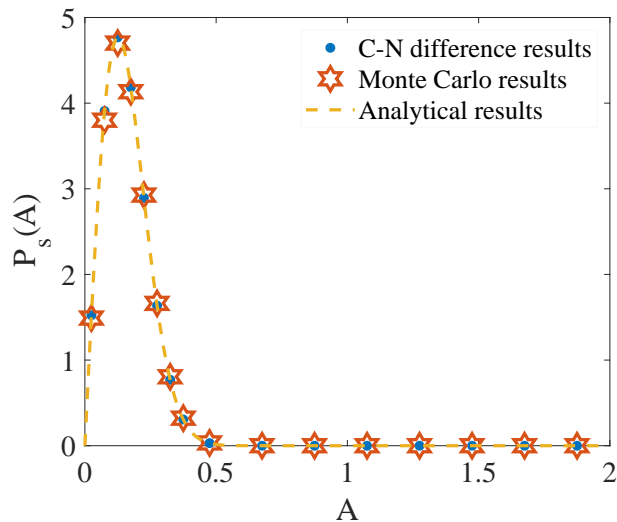}
	\caption{The SPDF $P_s(A)$ of the amplitude $A$ with additive colored noise for $\beta_1=8$, $\beta_2=2$, $\omega_{0}=120\times 2\pi$, $D_1=5\times10^5$, $\tau_1=0.003$, $v_0=-9$. The dashed line and the dot represent the analytical and numerical results of the FPK equation (\ref{FPK}), respectively. Hexagon denotes the Monte Carlo results of the original system (\ref{main eq}).}
	\label{fig1}
\end{figure}

We first consider that the thermoacoustic system (\ref{main eq}) is only excited by additive colored noise, that is, $D_1>0$, $D_2=0$. The dynamical behavior of stochastic systems can be derived from the PDF. According to the analysis in Section \ref{sec2}, we yield the approximate analytic solution of the SPDF. Figure \ref{fig1} displays the SPDF $P_s(A)$ of the amplitude $A$ obtained by three ways. The results obtained by Monte Carlo simulations based on the dynamical equation (\ref{main eq}) are in agreement with the analytical ones, which shows the validity of the stochastic averaging method. In addition, the results of the Crank-Nicolson (C-N) difference method based on the FPK equation (\ref{FPK}) are consistent with those of the two methods mentioned above, which shows the accuracy of the numerical methods.

\begin{figure}[htbp]
	\centering
	\includegraphics[width=8cm]{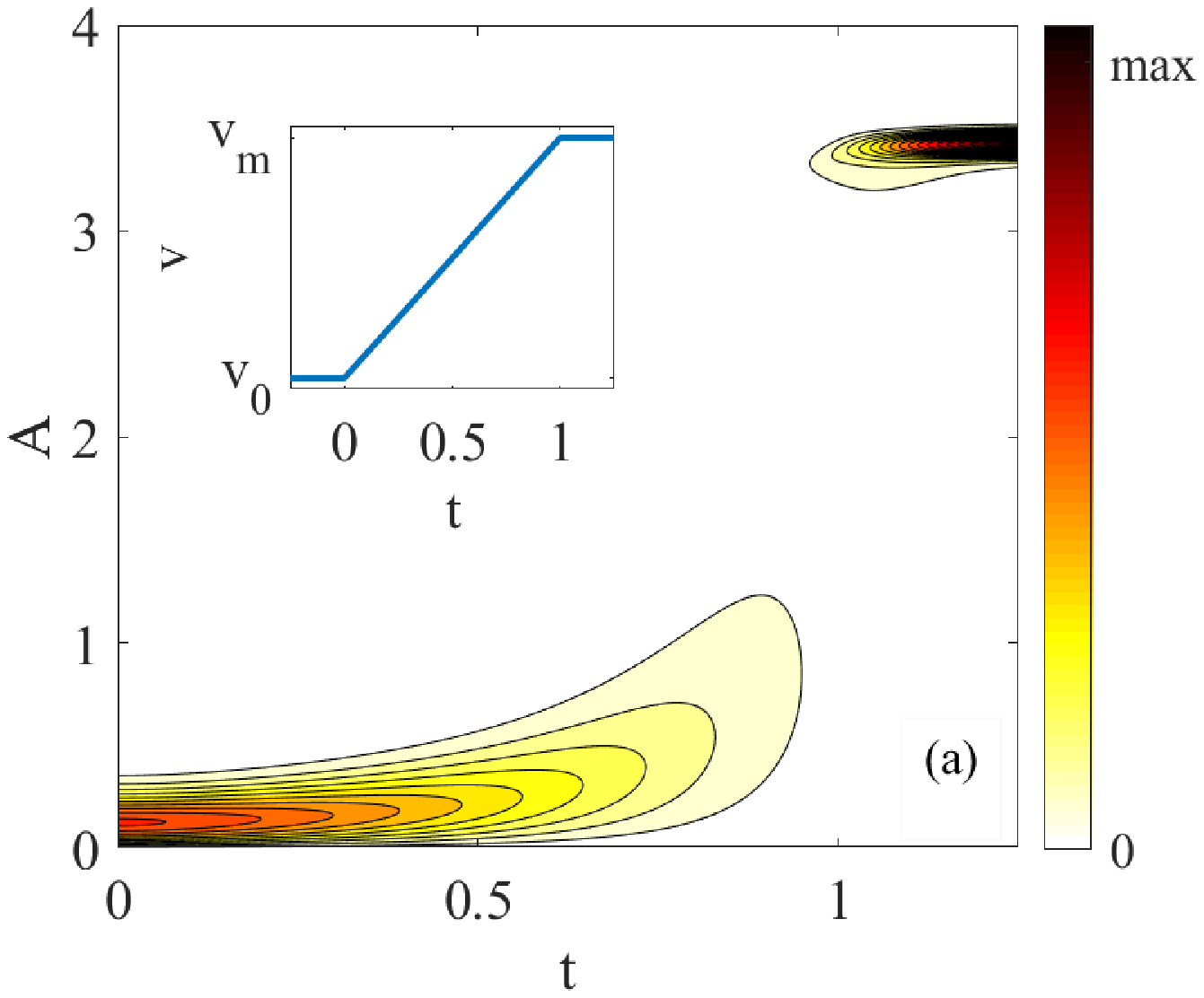}
	\includegraphics[width=8cm]{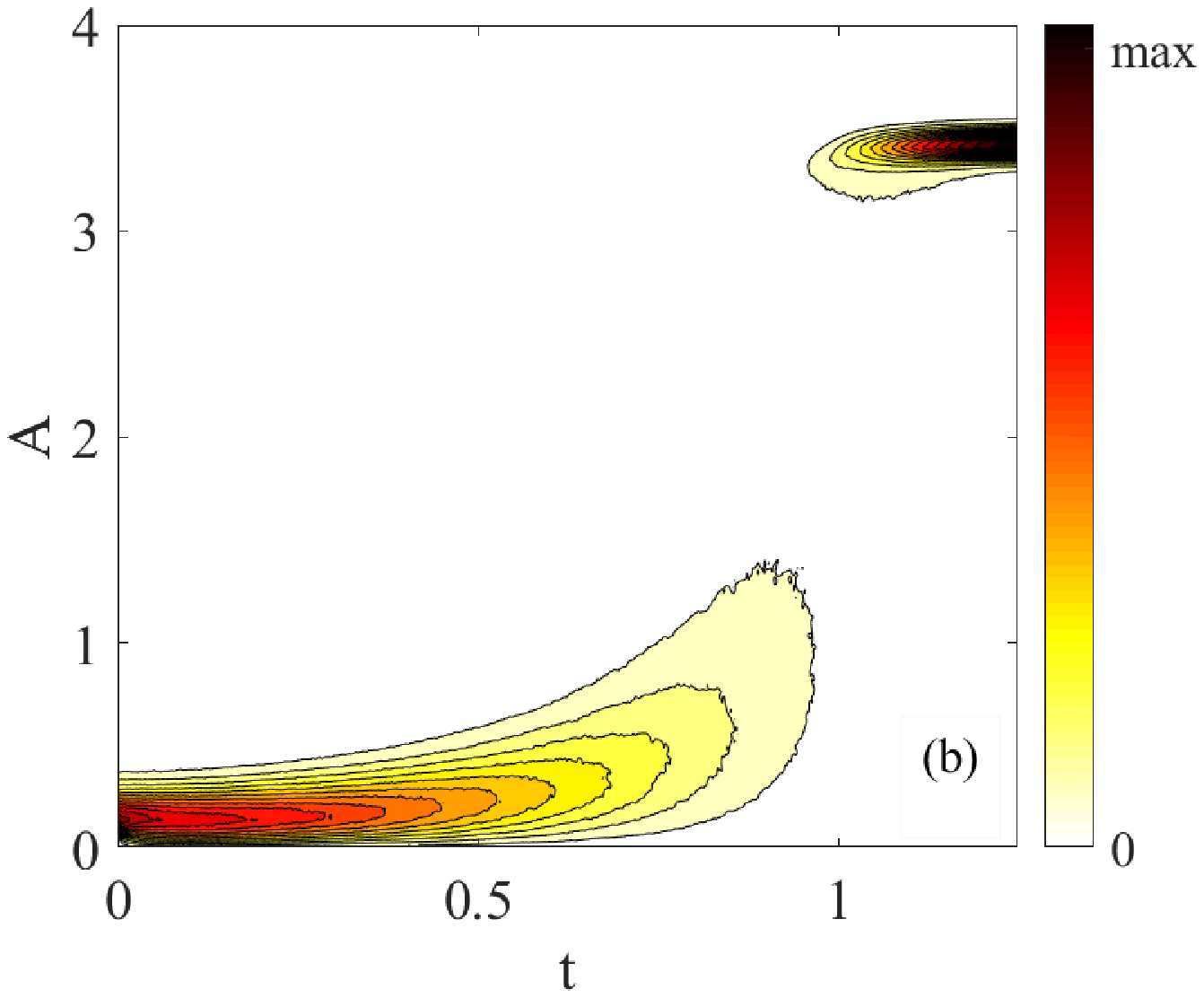}
	\caption{The contour plot represents the PDF $P(A,t)$ of the system (\ref{main eq}) with the time-varying parameter $v(t)$ under the additive colored noise. The internal graph is the curve of $v(t)=v_0+Rt$ for $v_0=-9$, $R=20$ and the changing time $t_c=1$. Other parameters are $\beta_1=8$, $\beta_2=2$, $\omega_{0}=120\times 2\pi$, $D_1=5\times10^5$, $\tau_1=0.003$. (a) C-N difference results of the FPK equation (\ref{FPK}); (b) Monte Carlo results of the original dynamical system (\ref{main eq}).}
	\label{fig2}
\end{figure}

Compared with the steady state, it is difficult to obtain the analytical solution for the transient dynamical behavior. But it can be discussed by solving the reduced FPK equation numerically. Figure \ref{fig2} illustrates the consistency of the two numerical methods. It presents the PDF $P(A,t)$ of the amplitude $A$ , when the parameter function $v(t)$ increases linearly from $v_0$ to $v_m$ with a rate $R=20$. The initial distribution is the SPDF $P_s(A)$ at $v_0={-9}$. Figure \ref{fig2}(a) indicates the C-N difference results with $\beta_1=8$, $\beta_2=2$, $\omega_{0}=120\times 2\pi$, $D_1=5\times10^5$, $\tau_1=0.003$, $v_0=-9$ , which is in good agreement with the Monte Carlo results from the equation (\ref{main eq}) in Figure \ref{fig2}(b). We mainly use the difference method to study the follow-up content.

\begin{figure}[htbp]
	\centering
	\includegraphics[width=12cm]{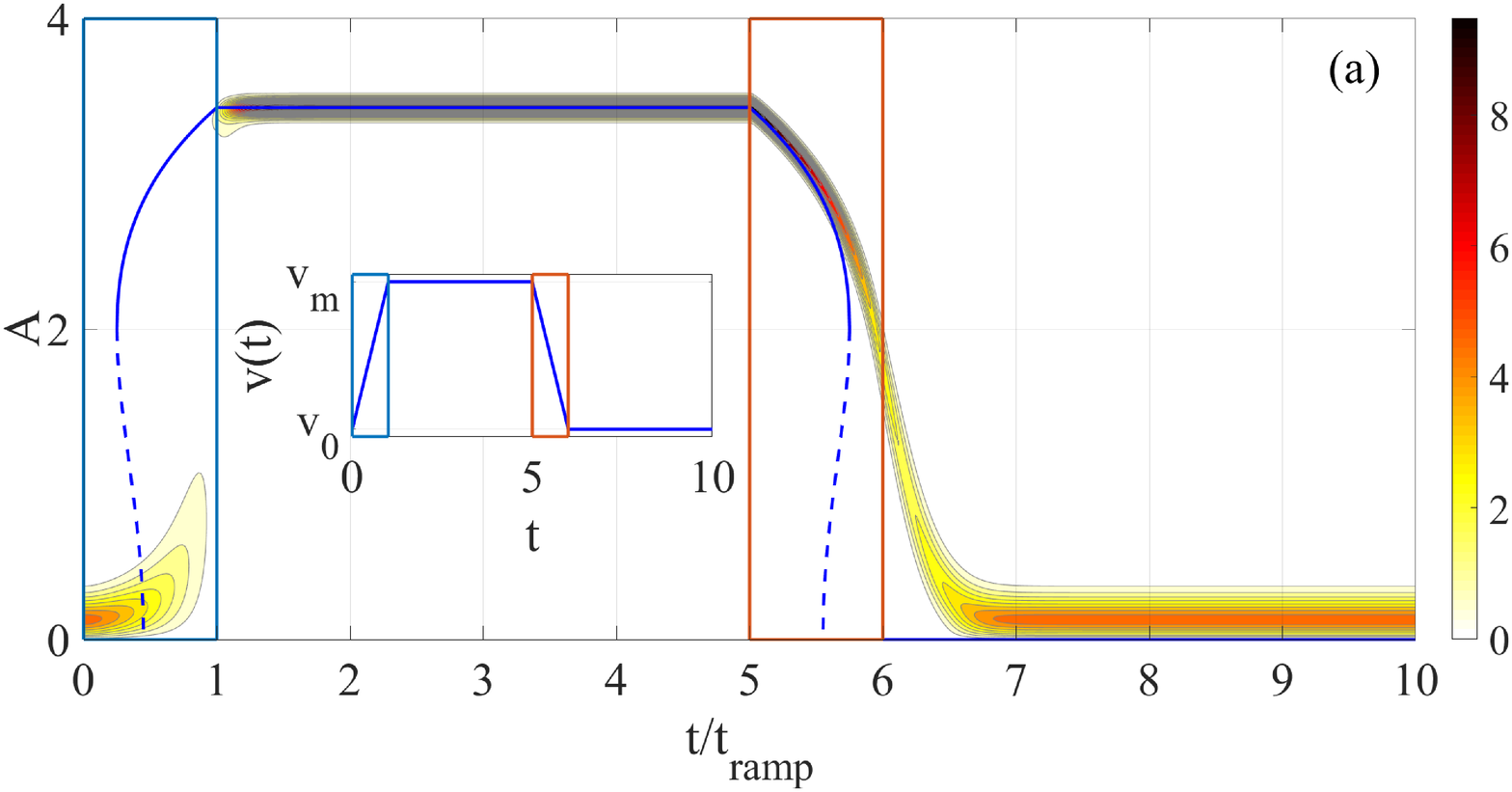}
	\includegraphics[width=12cm]{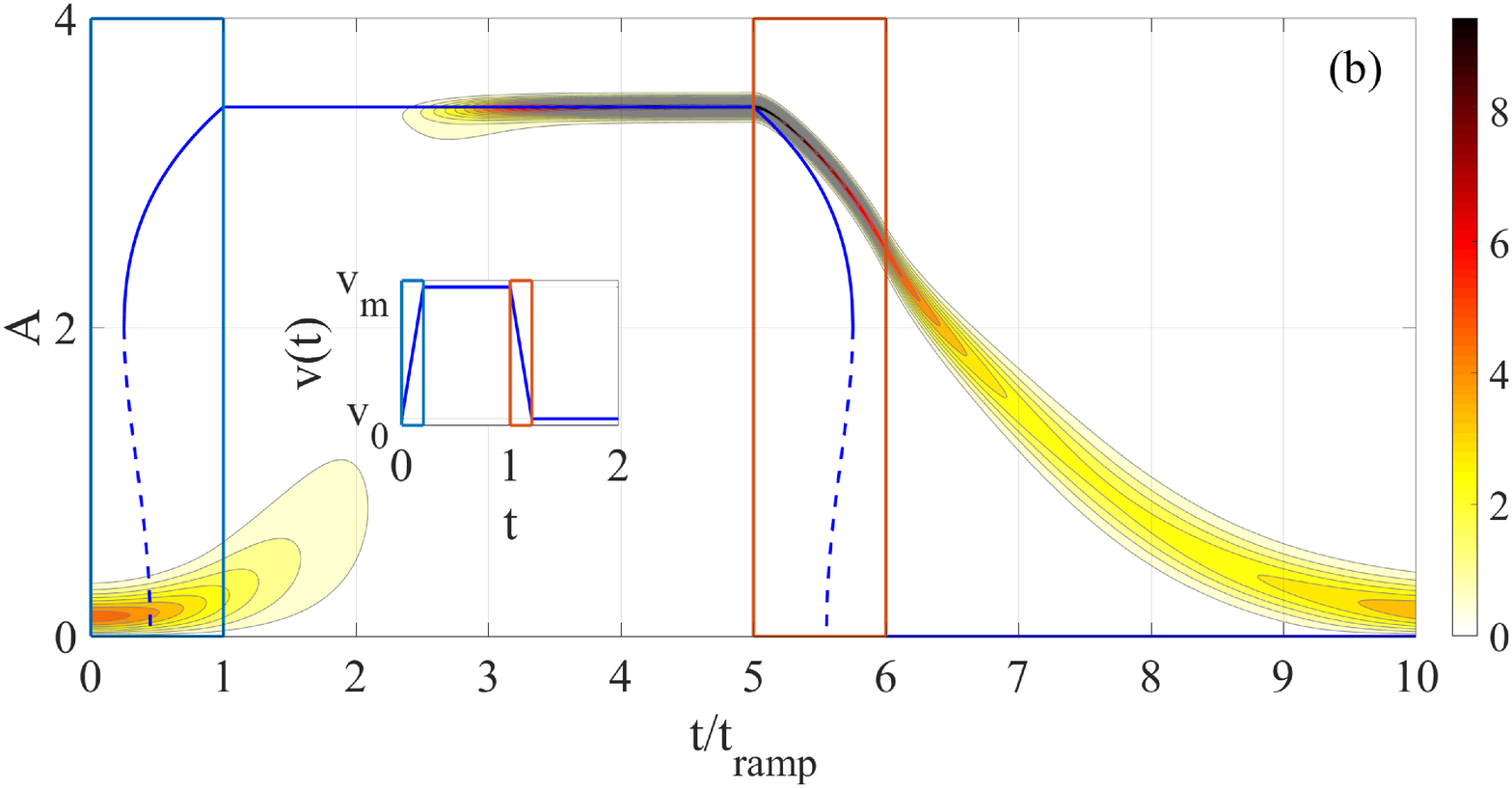}
	\caption{The effect of different rates $R$ of the time-varying parameter $v(t)$ with fixed $v_0=-9$ and $v_m=11$ on transient dynamical behavior. The internal figure shows the variation of $v(t)$ with time. The blue line is the quasi-steady deterministic results. Abscissa is the time after normalization.  Other parameters are $\beta_1=8$, $\beta_2=2$, $\omega_{0}=120\times 2\pi$, $D_1=5\times10^5$, $\tau_1=0.003$. (a) $R=20$; (b) $R=100$.}
	\label{fig3}
\end{figure}

Now we consider that $v(t)$ increases linearly from the initial value $v_0$ to $v_m$ but then decreases linearly from $v_m$ to $v_0$ after a period. When the correlation time $\tau_1$ approaches to zero, the system becomes a white noise excitation, which was discussed in [17]. The main difference between the cases of colored and white noise is that the probability of the particles between two stable sets is very small for colored noise, and there is no line region with a certain probability (see Figure \ref{fig3}). This means that  colored noise makes the state of the system change more quickly and thoroughly than white noise does. However, it is worthy noting that this phenomenon is more obvious when  $v_0$ increases linearly to $v_m$, but does not occur when $v_m$ decreases to $v_0$. 

Figure \ref{fig3} represents the contour plot of the PDF $P(A,t)$ with $v(t)$ which is shown by the internal diagram. To facilitate the comparison of the results for different $R$, the abscissa use $t=t/t_{ramp}$ and $t_{ramp}=(v_m-v_0)/R$ to normalize the time. The corresponding stationary bifurcation diagram with the parameter $v$ for the deterministic system in blue is used as a reference. In the quasi-steady deterministic case, the time for the system to end the bistable state is 0.45, then the system becomes monostable again. When the rate is introduced into the parameters and the system is excited by colored noise, the time of a complete transition from the low-amplitude state to the high-amplitude one is about one, as shown in Figure \ref{fig3}(a). This implies that the rate-dependent tipping-delay phenomenon occurs compared with the quasi-steady picture of the system.

Furthermore, we uncover that the faster the parameter changes, the more obvious the tipping-delay is. It is important to point out that, ‘more obvious tipping-delay’ does not mean that tipping happened late in real time. In fact, we unify the time scale and keep the values of $v_0$ and $v_m$ unchanged. Thus, the quasi-steady deterministic results which are the references in the Figures \ref{fig3}(a) and \ref{fig3}(b) are the same. When the rate $R$ increases, the normalized time at the moment of the tipping increases, too. This implies that the value of $v$ at which tipping occurs is increasing. But in real time, the tipping  occurs faster. 

\subsection{The case of multiplicative colored noise}
\begin{figure}[htbp]
	\centering
	\includegraphics[width=8cm]{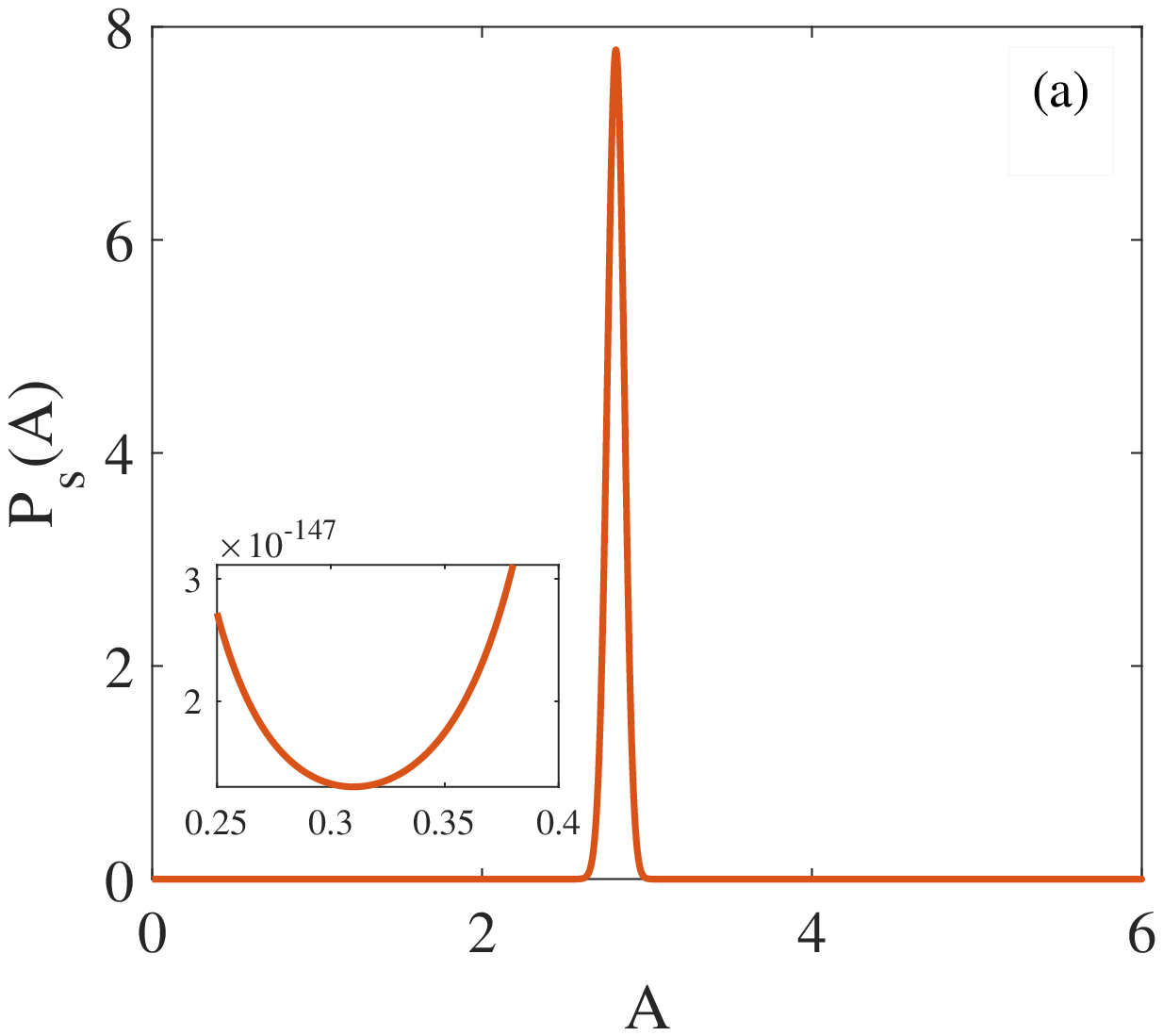}
	\includegraphics[width=8cm]{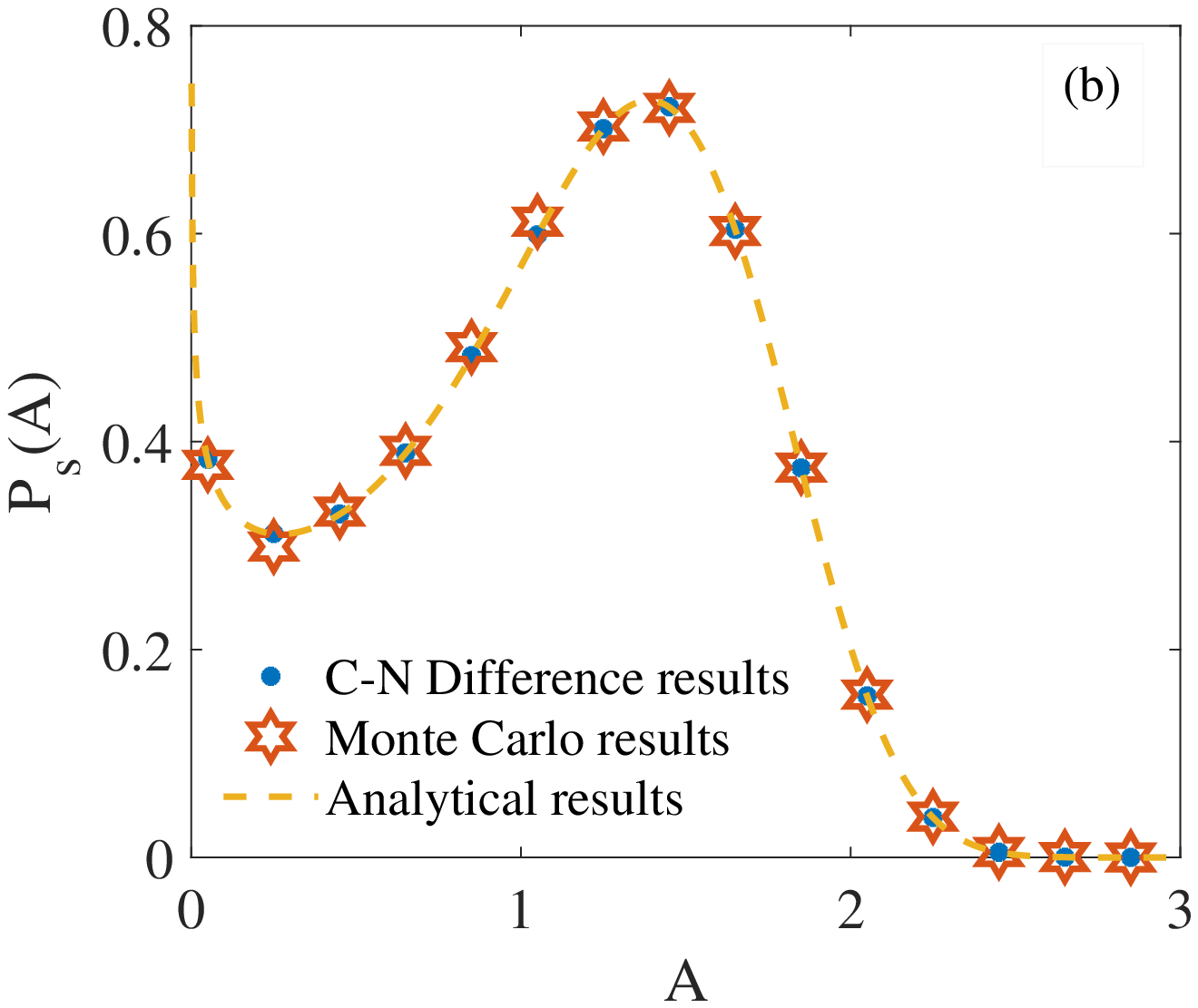}
	\caption{The SPDF $P_s(A)$ of the amplitude $A$ with multiplicative colored noise for $\omega_{0}=120\times 2\pi$, $D_2=5\times10^5$, $\tau_2=0.003$. (a) $\beta_1=8$, $\beta_2=2$, $v_0=-0.2$; (b) $\beta_1=0.1$,  $\beta_2=0.1$, $v_0=-0.012$.}
	\label{fig4}
\end{figure}

\begin{figure}[htbp]
	\centering
	\includegraphics[width=8cm]{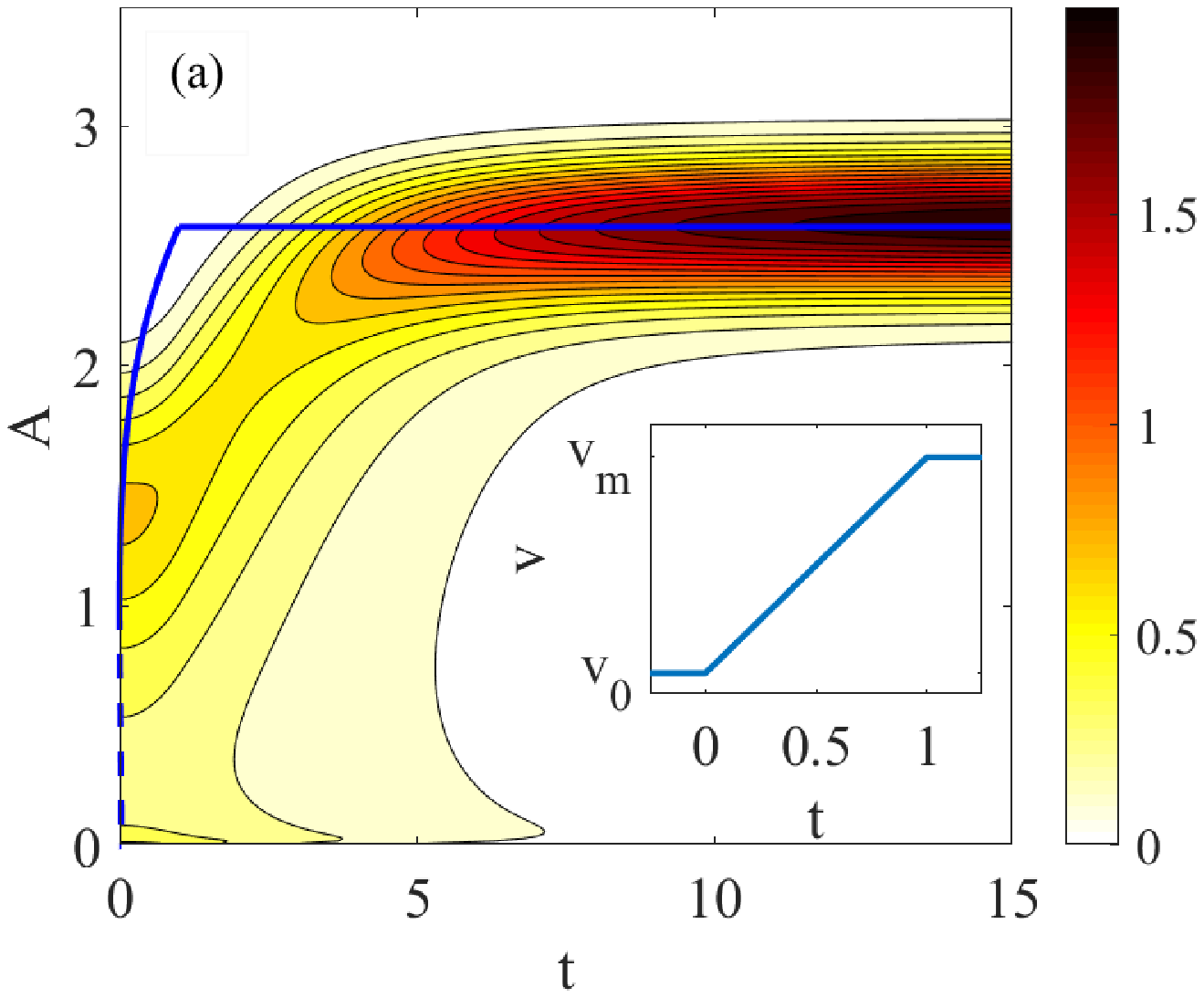}
	\includegraphics[width=8cm]{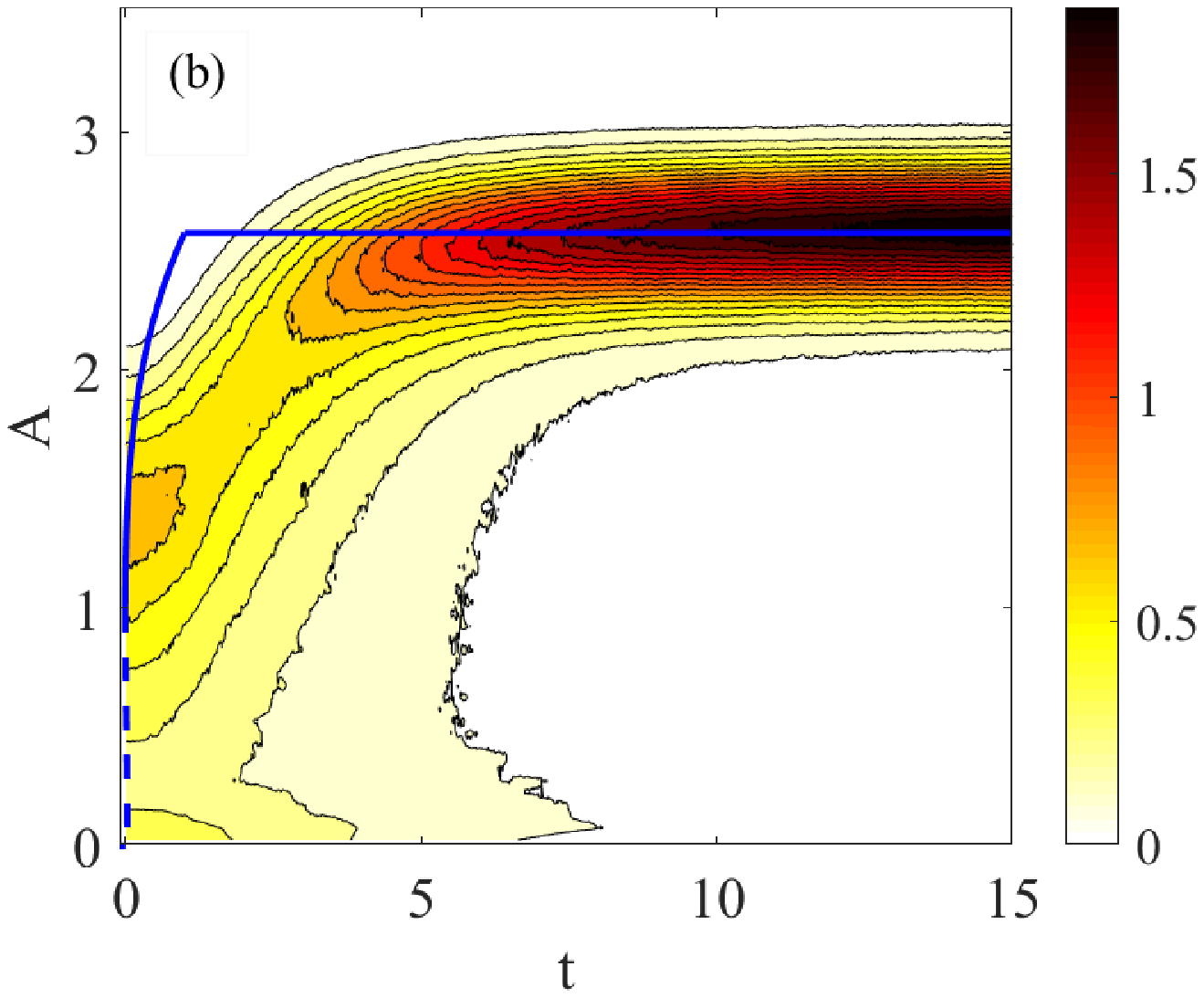}
	\caption{The contour plot represents the PDF $P(A,t)$ of the system with time-varying parameters under multiplicative colored noise. The internal graph is the curve of $v(t)=v_0+Rt$ for $v_0=-0.012$, $R=0.4$. Other parameters are $\beta_1=0.1$,  $\beta_2=0.1$, $\omega_{0}=120\times 2\pi$, $D_2=5\times10^5$, $\tau_2=0.003$. (a) C-N difference numerical results of the FPK equation (\ref{FPK}); (b) Monte Carlo simulations of the original dynamical system (\ref{main eq}).}
	\label{fig5}
\end{figure}

When $D_1=0$, and $D_2>0$, the thermoacoustic system (\ref{main eq}) specifies to be excited only by multiplicative colored noise. The analytic solution of the SPDF can be obtained after satisfying the existence condition. When we follow the parameter values under additive colored noise, the SPDF has two extremums, but the first extremum is not obvious, as shown in the Figure \ref{fig4}(a). This makes it meaningless to study tipping behavior later, so we consider new parameters as $\omega_{0}=120\times 2\pi$, $D_2=5\times10^5$, $\tau_2=0.003$, $\beta_1=0.1$,  $\beta_2=0.1$, $v_0=-0.012$. A consistency of the stationary solution obtained by the C-N difference, the Monte Carlo and analytical methods is shown in the Figure \ref{fig4}(b). Considering that $v(t)$ varies linearly with time, Figure $\ref{fig5}$ shows the transient PDFs of the amplitude $A$ from the FPK equation (\ref{FPK}) and the dynamical system (\ref{main eq}) respectively, which indicates that the stochastic averaging method is applicable to the case of only multiplicative noise excitation.

\begin{figure}[htbp]
	\centering
	\includegraphics[width=8cm]{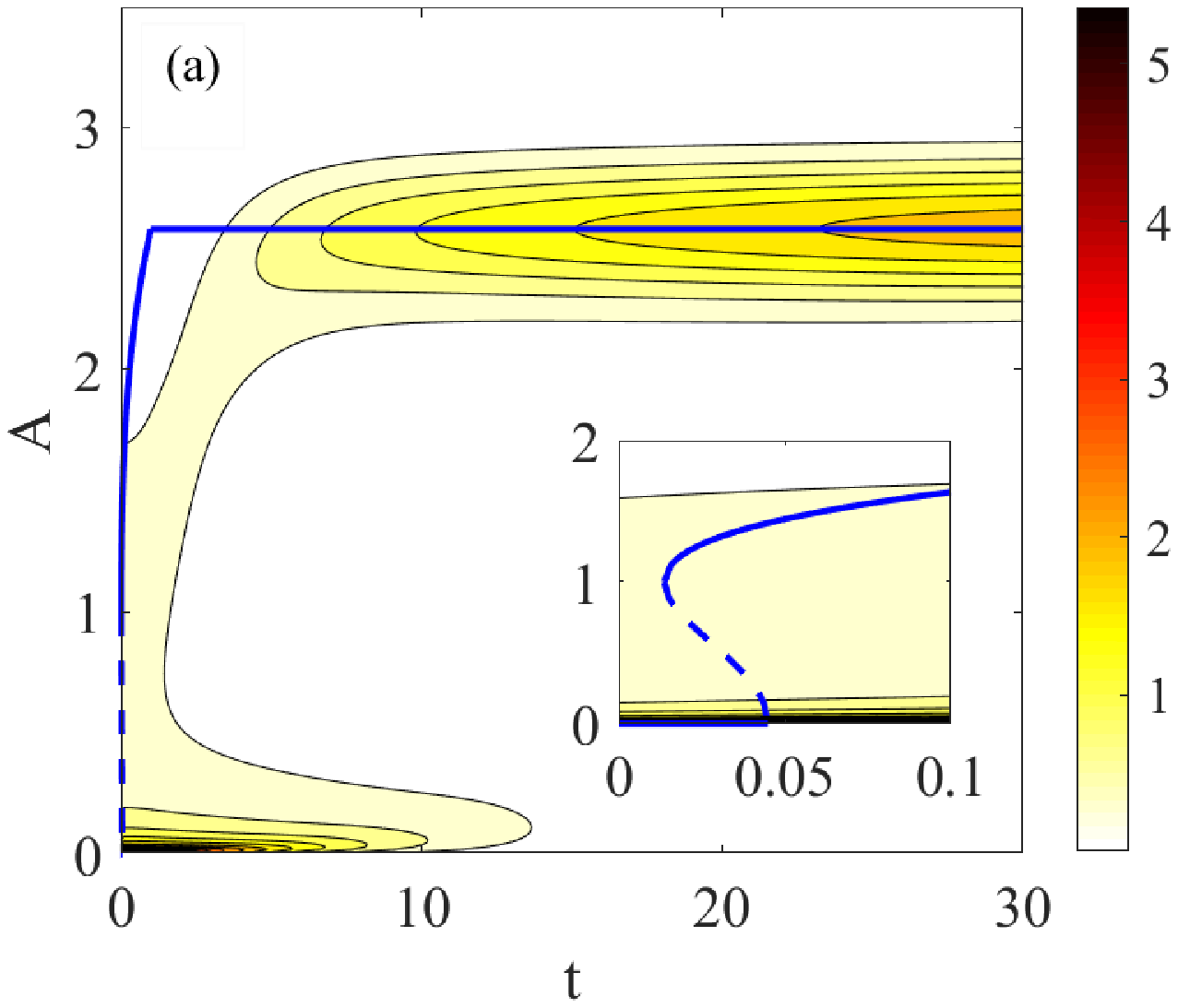}
	\includegraphics[width=8cm]{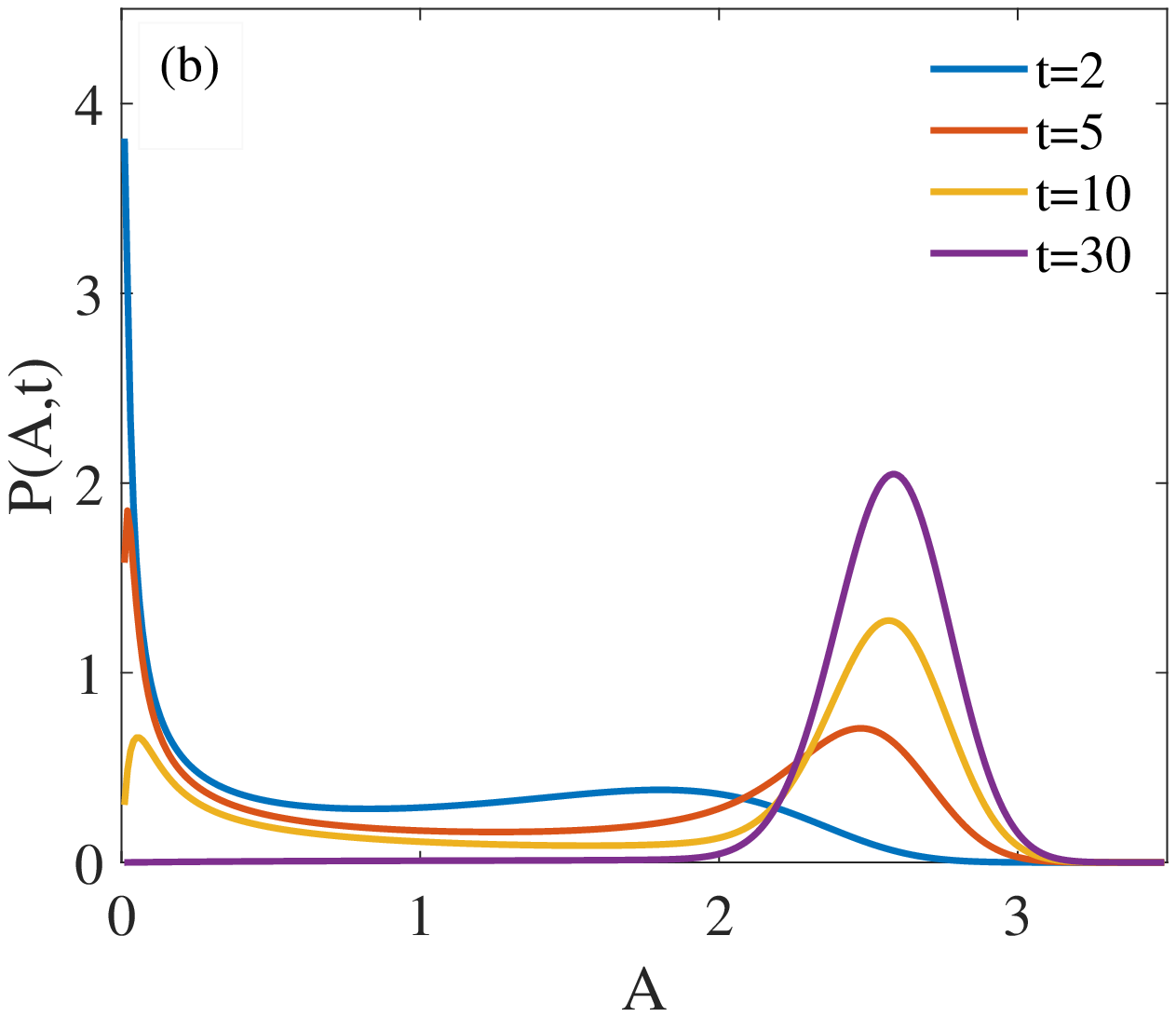}
	\includegraphics[width=8cm]{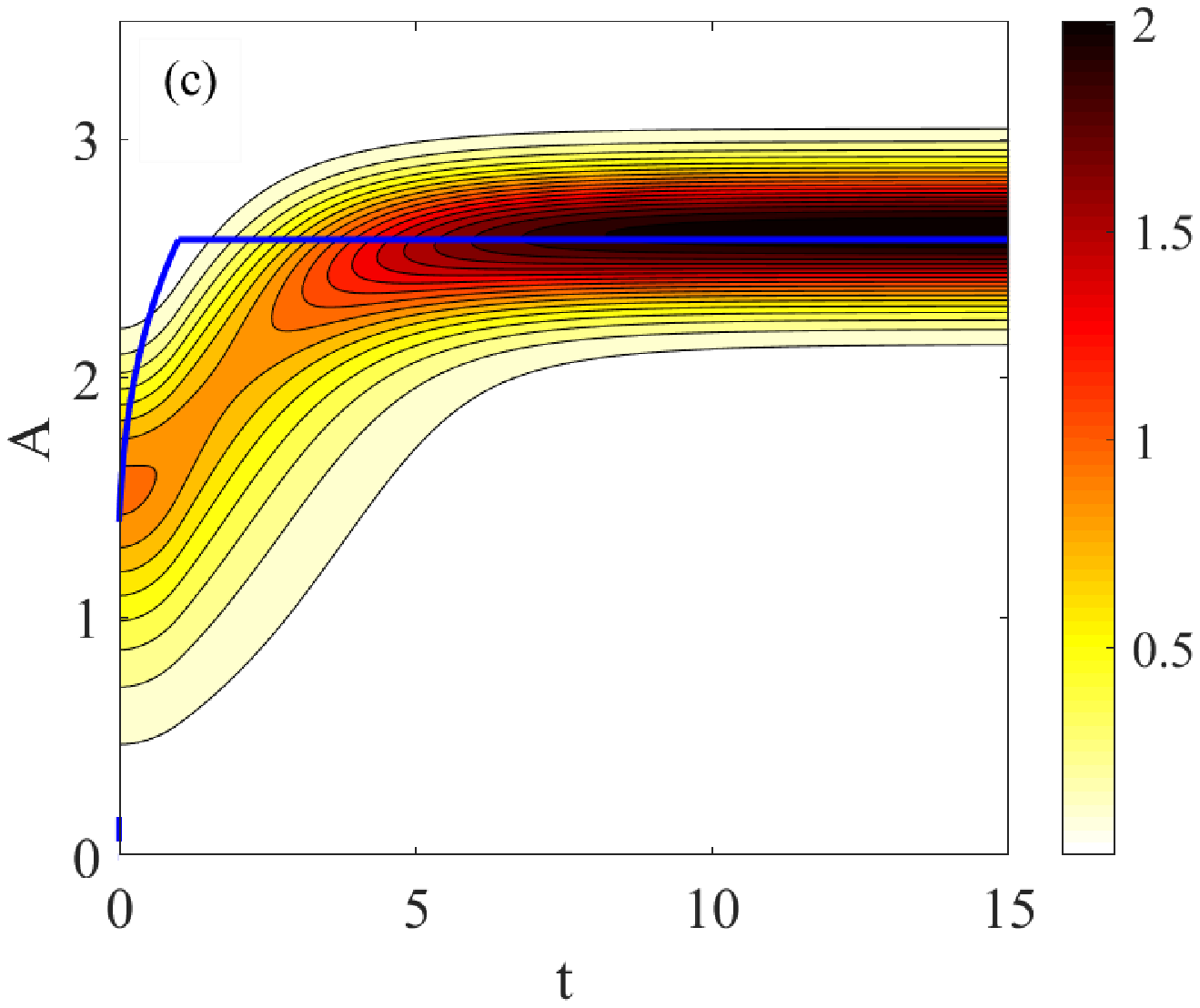}
	\includegraphics[width=8cm]{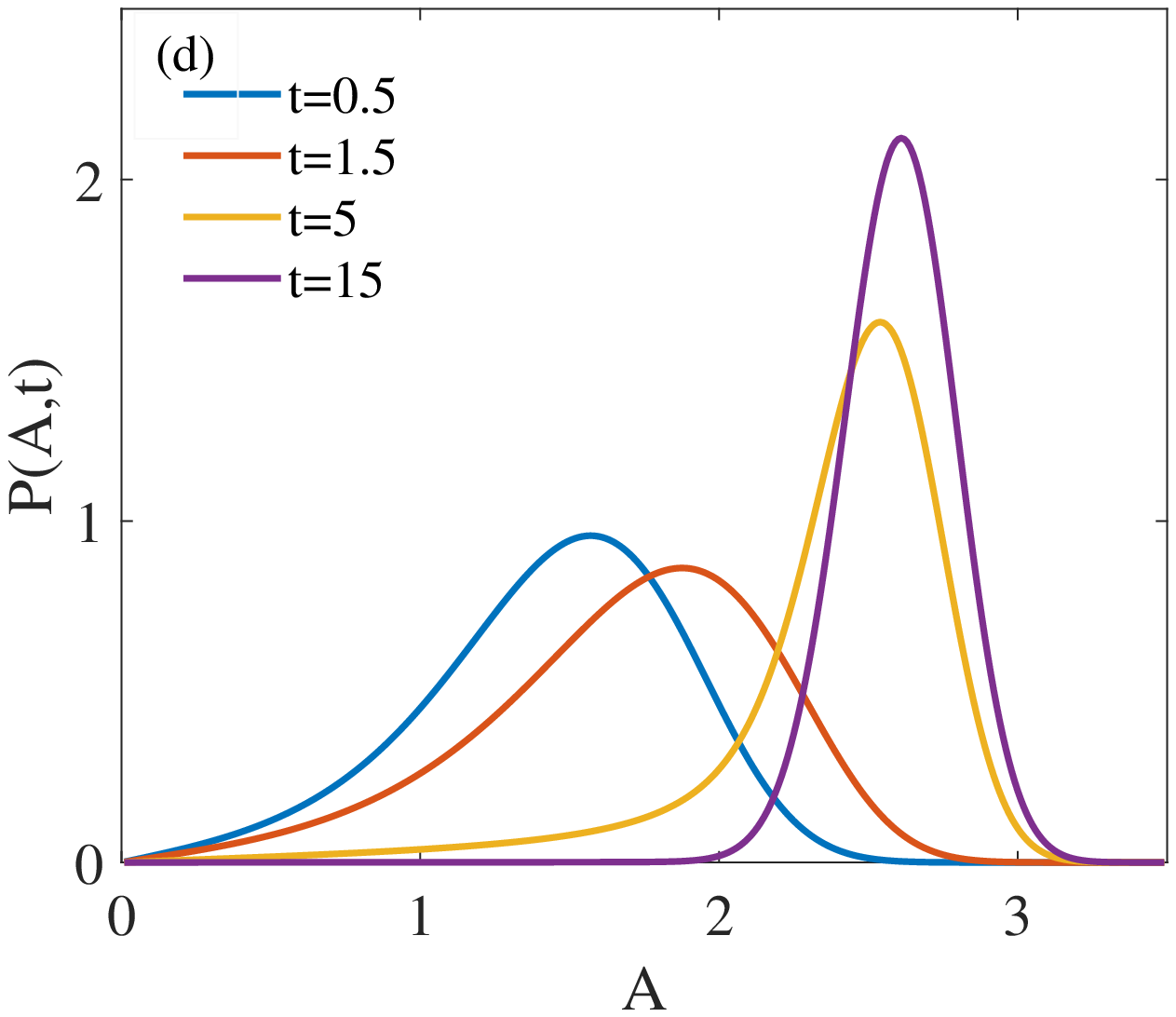}
	\caption{The effect of different initial values $v_0$ of the time-varying parameter $v(t)$ for $R=0.4$, $t_c=1$ on the transient dynamical behavior. (b) and (d) are two-dimensional images obtained by selecting several time points in (a) and (c), respectively. Other parameters are $\beta_1=0.1$,  $\beta_2=0.1$, $\omega_{0}=120\times 2\pi$, $D_2=5\times10^5$, $\tau_2=0.003$. (a,b) $v_0=-0.018$; (c,d) $v_0=0$.}
	\label{fig6}
\end{figure}

Next, we study the effects of the initial value $v_0$, the ramp rate $R$ and the changing time $t_c$  on the stochastic dynamical behaviors of the thermoacoustic system. Figure \ref{fig6} displays the results with the initial values $v_0=-0.018$ and $v_0=0$. As shown in Figure \ref{fig6}(a), the bistability ends at $t=0.045$ in the deterministic case, while the rate-dependent system with multiplicative colored noise ends at about $t=15$, implying that the rate-dependent tipping-delay phenomenon occurs. In addition, the delay time of the rate-dependent tipping-delay becomes longer when the initial value reduces in a certain range.

From Figure \ref{fig6}(c), the PDF is always in a unimodal state from the beginning, and there is no switching between the bimodal and unimodal region. The system enters the high-amplitude state completely after $t=1$ in the deterministic case, while after $t=5$ for the rate-dependent system with multiplicative colored noise. In other words, the system still has a delay phenomenon. It still confirms the above result that the smaller the initial value, the longer the delay time of the rate-dependent tipping-delay phenomenon.

\begin{figure}[htbp]
	\centering
	\includegraphics[width=8cm]{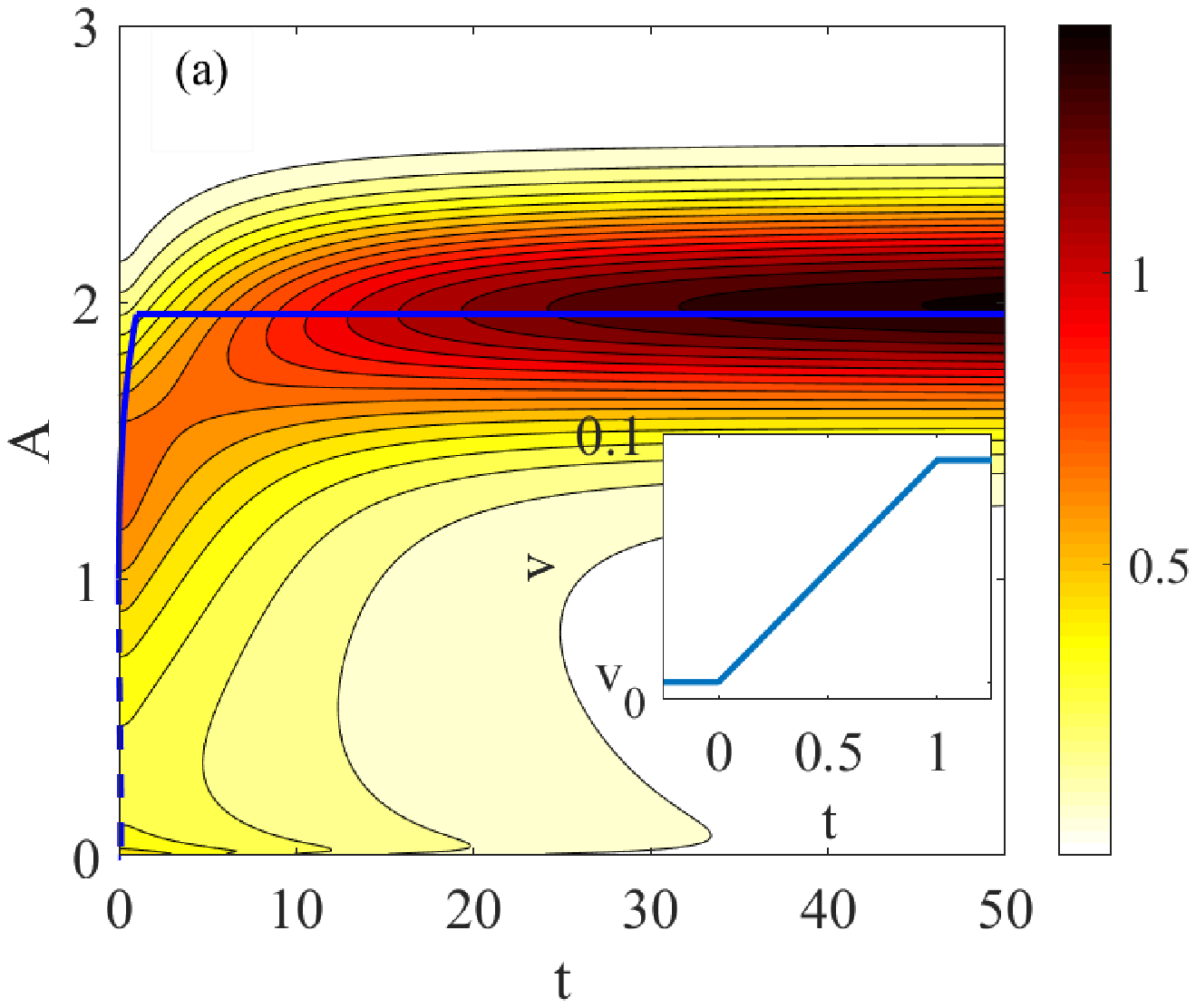}
	\includegraphics[width=8cm]{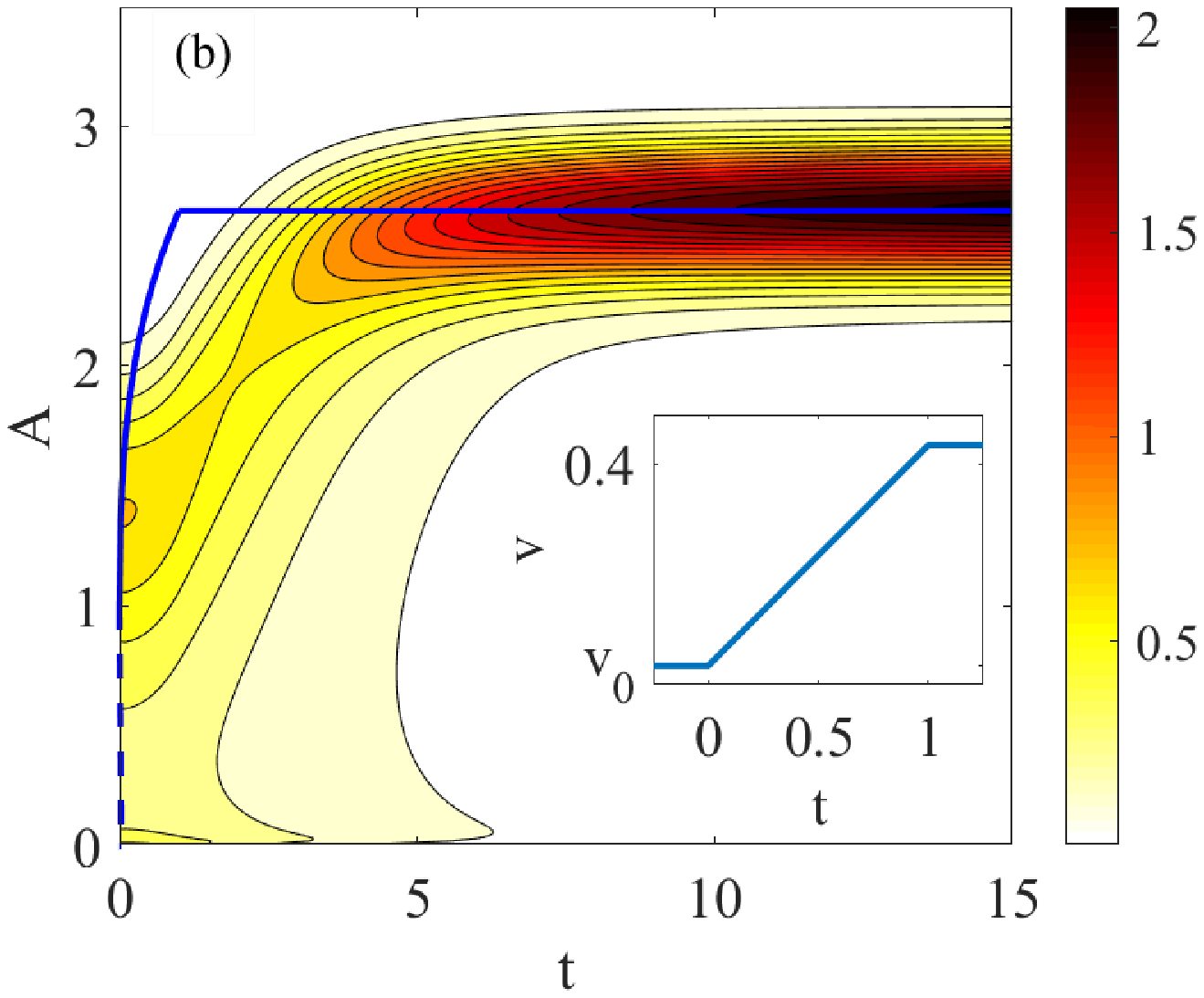}
	\caption{The effect of different rate $R$ of the time-varying parameter $v(t)$ for $v_0=-0.012$, $t_c=1$ on the transient dynamical behavior. Other parameters are $\beta_1=0.1$,  $\beta_2=0.1$, $\omega_{0}=120\times 2\pi$, $D_2=5\times10^5$, $\tau_2=0.003$. (a) $R=0.1$; (b) $R=0.45$.}
	\label{fig7}
\end{figure}

We consider now the influence of the ramp rate on the transient dynamical behavior of the system under the excitation of multiplicative colored noise. As can be seen from Figure \ref{fig7}, the delay time of the rate-dependent tipping-delay phenomenon decreases with the increase of the rate. Note that this is different from the phenomenon we discussed above under the excitation of additive colored noise. In fact, the aspects discussed here are also different from additive noise. In the additive case, we keep the initial and final values of $v$ unchanged, so for different rates, the changing time of $v$ is different. While in the multiplicative case, we keep the initial value and changing time consistent, and different rate makes the final value different. If we discuss it in the same way, we will get a conclusion similar to the additive case. Both of the above strategies have a clear practical significance, and choosing which one depends on the actual thermoacoustic system requirements.

\begin{figure}[htbp]
	\centering
	\includegraphics[width=8cm]{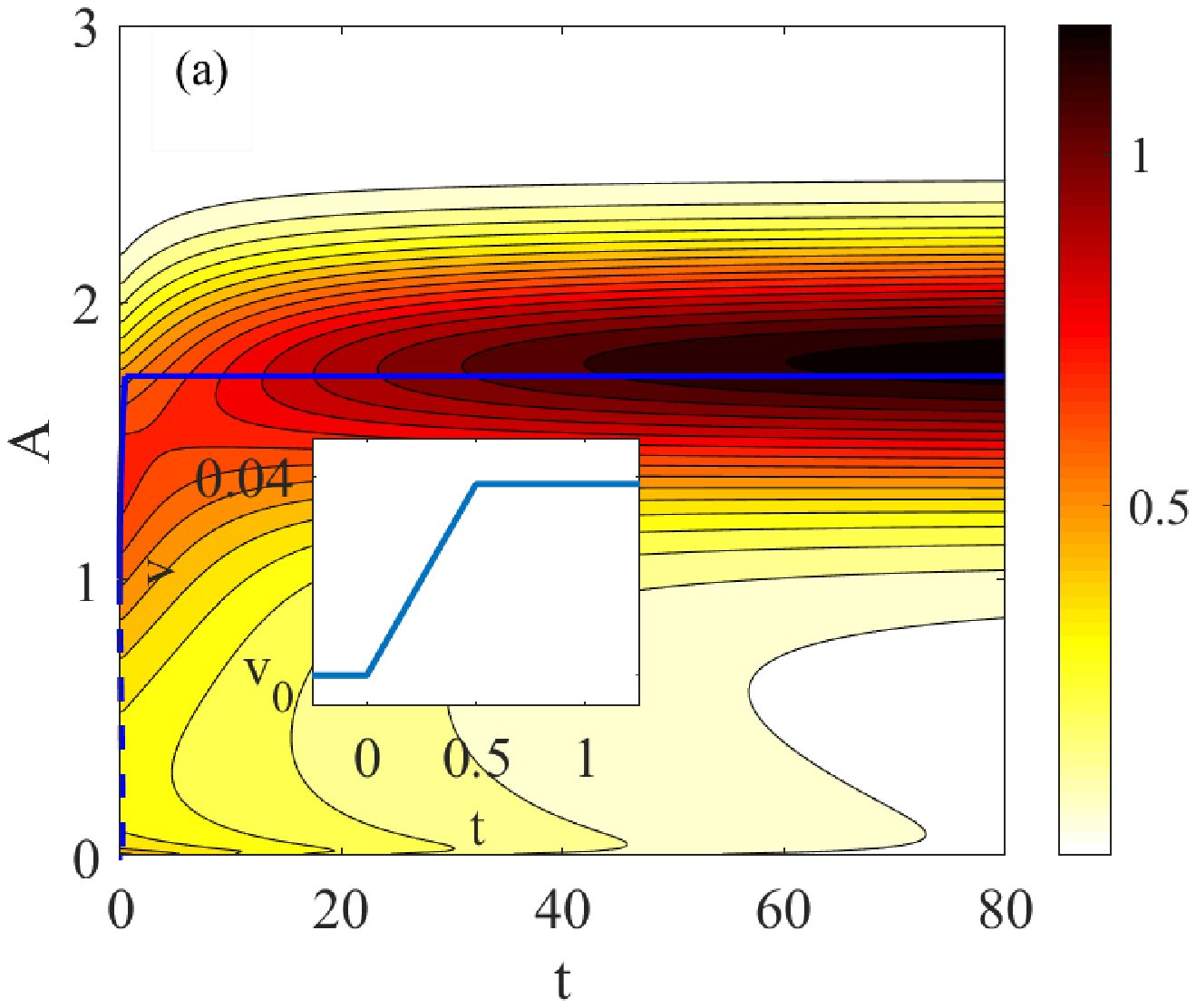}
	\includegraphics[width=8cm]{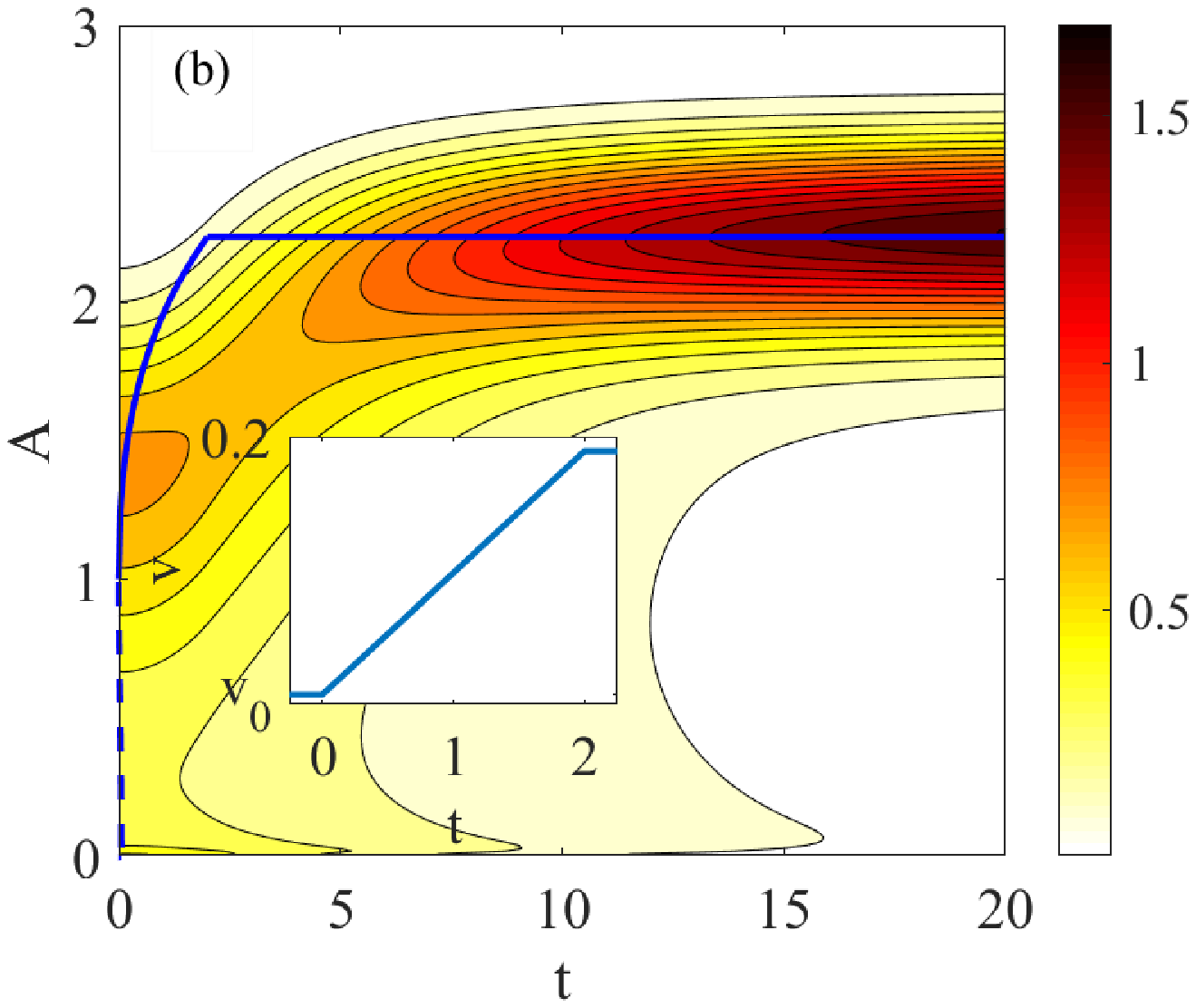}
	\caption{The effect of different changing time $t_c$ of the time-varying parameter $v(t)$ for $v_0=-0.012$, $R=0.1$ on the transient dynamical behavior. Other parameters are $\beta_1=0.1$,  $\beta_2=0.1$, $\omega_{0}=120\times 2\pi$, $D_2=5\times10^5$, $\tau_2=0.003$. (a) $t_c=0.5$; (b) $t_c=2$.}
	\label{fig8}
\end{figure}

Furthermore, we study the effect of parameter changing time $t_c$ on the transient dynamical behavior. Figure \ref{fig8} shows that with the increase of the parameter changing time, the time ending bimodality becomes smaller, thus, the shorter is the delay time of rate-dependent tipping-delay phenomenon. According to the form of $v(t)=v_0+Rt$, the longer the time of the parameter change is, the greater the value of $v$ finally arrives. Larger $v$ cancel out the partial inertia effect and shorten the delay time of the rate-dependent tipping-delay phenomenon.

\subsection{The case of additive and multiplicative colored noise}
\begin{figure}[htbp]
	\centering
	\includegraphics[width=9cm]{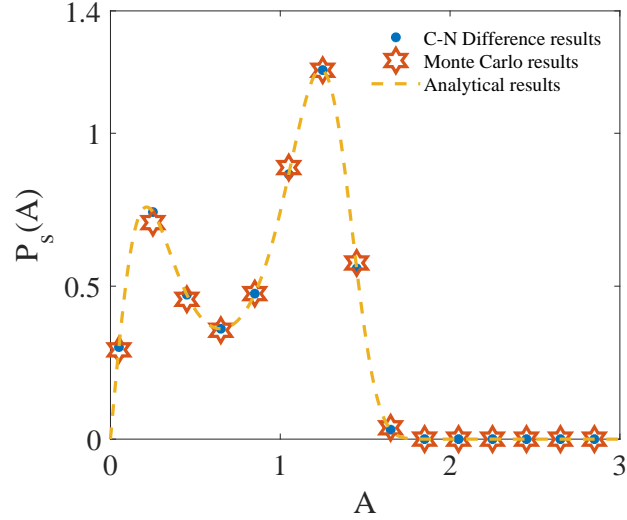}
	\caption{The SPDF $P_s(A)$ of the amplitude $A$ with multiplicative colored noise for $\beta_1=0.1$, $\beta_2=0.1$, $\omega_{0}=120\times 2\pi$, $D_1=6000$, $D_2=6000$, $\tau_1=0.007$, $\tau_2=0.001$, $v_0=-0.01$.}
	\label{fig9}
\end{figure}

\begin{figure}[htbp]
	\centering
	\includegraphics[width=8cm]{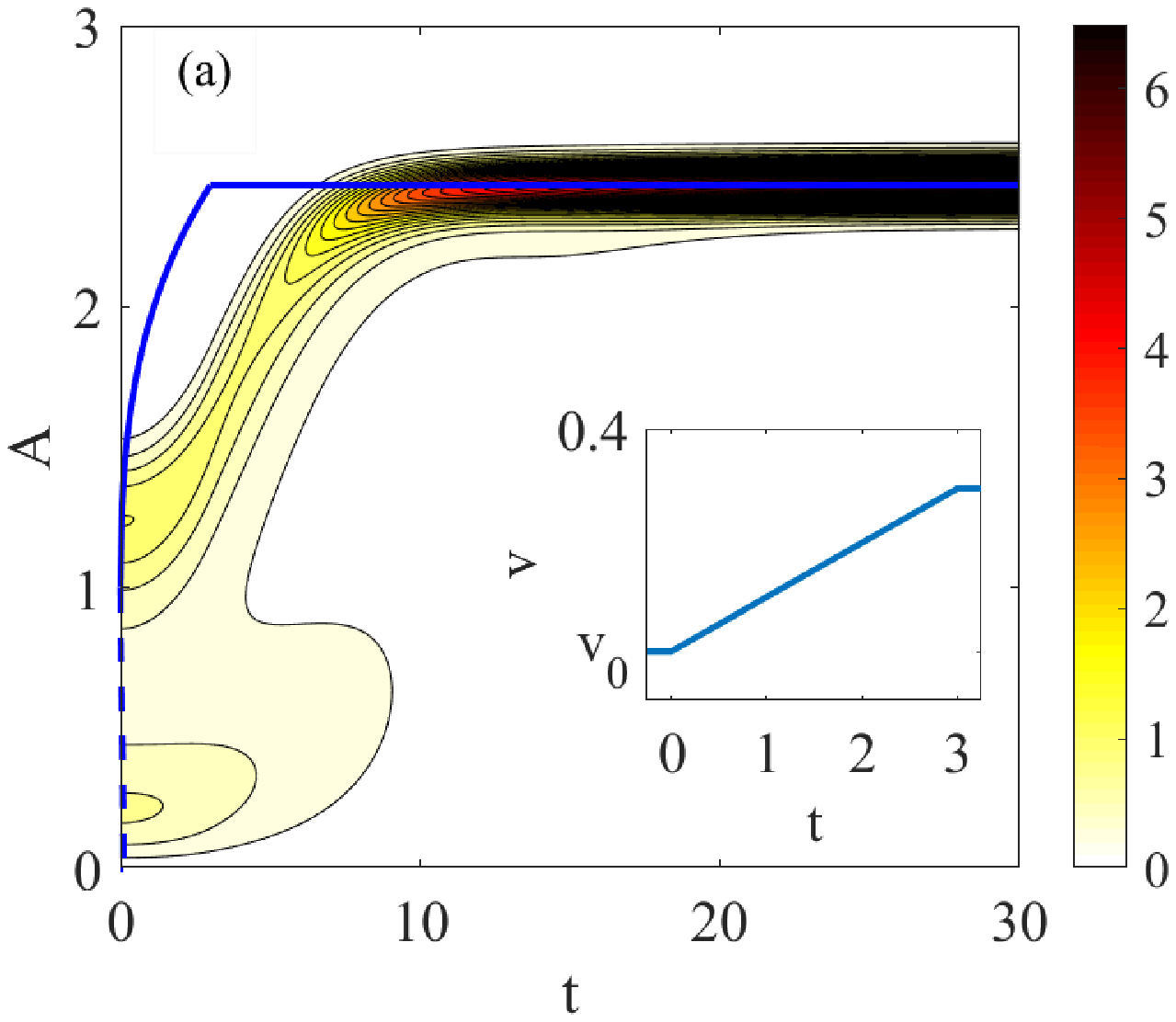}
	\includegraphics[width=8cm]{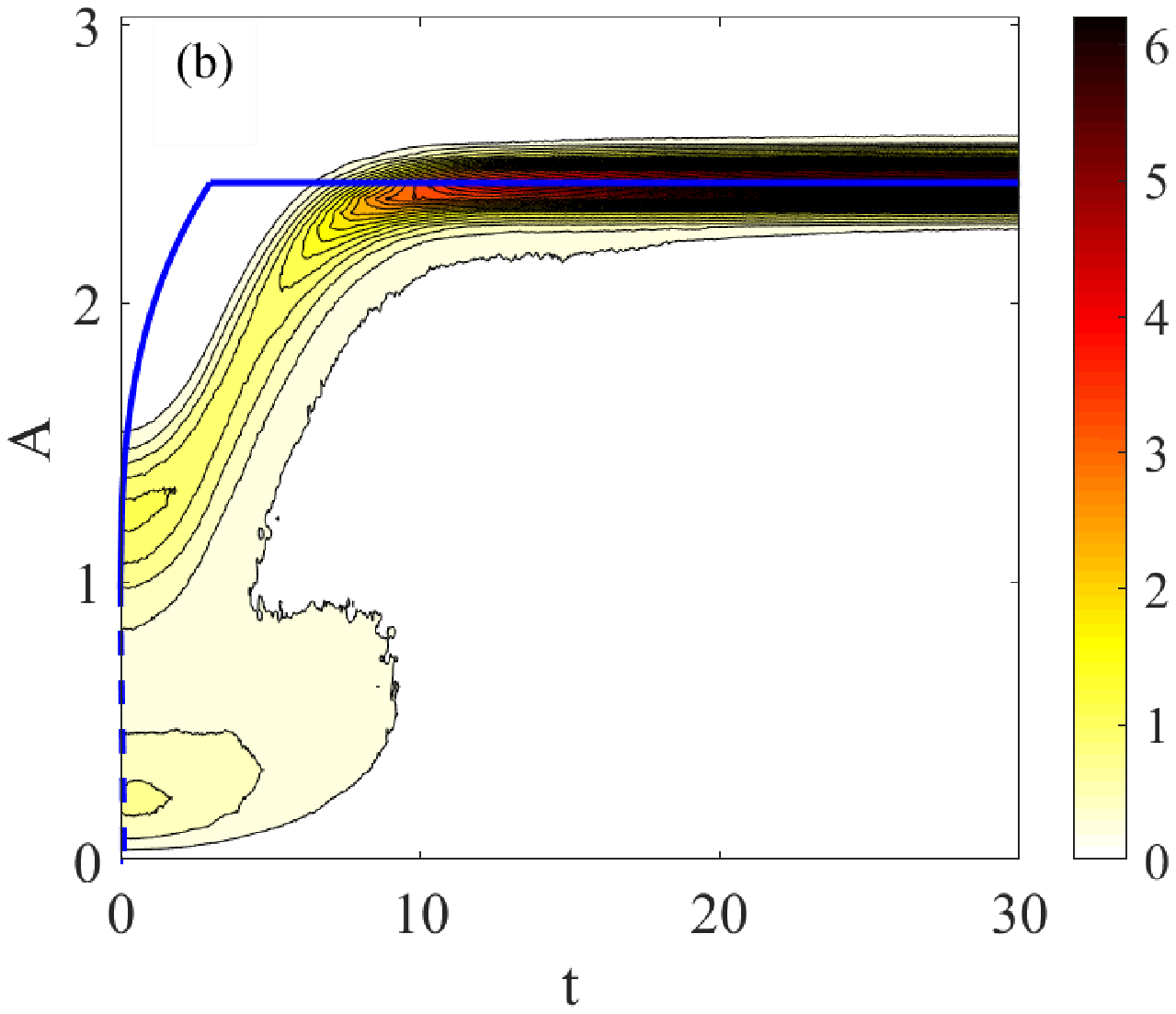}
	\caption{The contour plot represents the PDF $P(A,t)$ of the system with time-varying parameters under additive and multiplicative colored noise. The internal graph is the curve of $v(t)=v_0+Rt$ for $v_0=-0.01$, $R=0.1$ and $t_c=3$. Other parameters are $\beta_1=0.1$, $\beta_2=0.1$, $\omega_{0}=120\times 2\pi$, $D_1=6000$, $D_2=6000$, $\tau_1=0.007$, $\tau_2=0.001$. (a) C-N difference numerical results of the FPK equation (\ref{FPK}); (b) Monte Carlo simulations of the dynamical system (\ref{main eq}).}
	\label{fig10}
\end{figure}

When $D_1>0$ and $D_2>0$, the thermoacoustic system (\ref{main eq}) is excited by a combination of additive and multiplicative colored noises. Proceedings as in the above cases, the validity of the numerical methods and the applicability of the stochastic averaging method are verified in Figures \ref{fig9} and \ref{fig10}.

\begin{figure}[htbp]
	\centering
	\includegraphics[width=8cm]{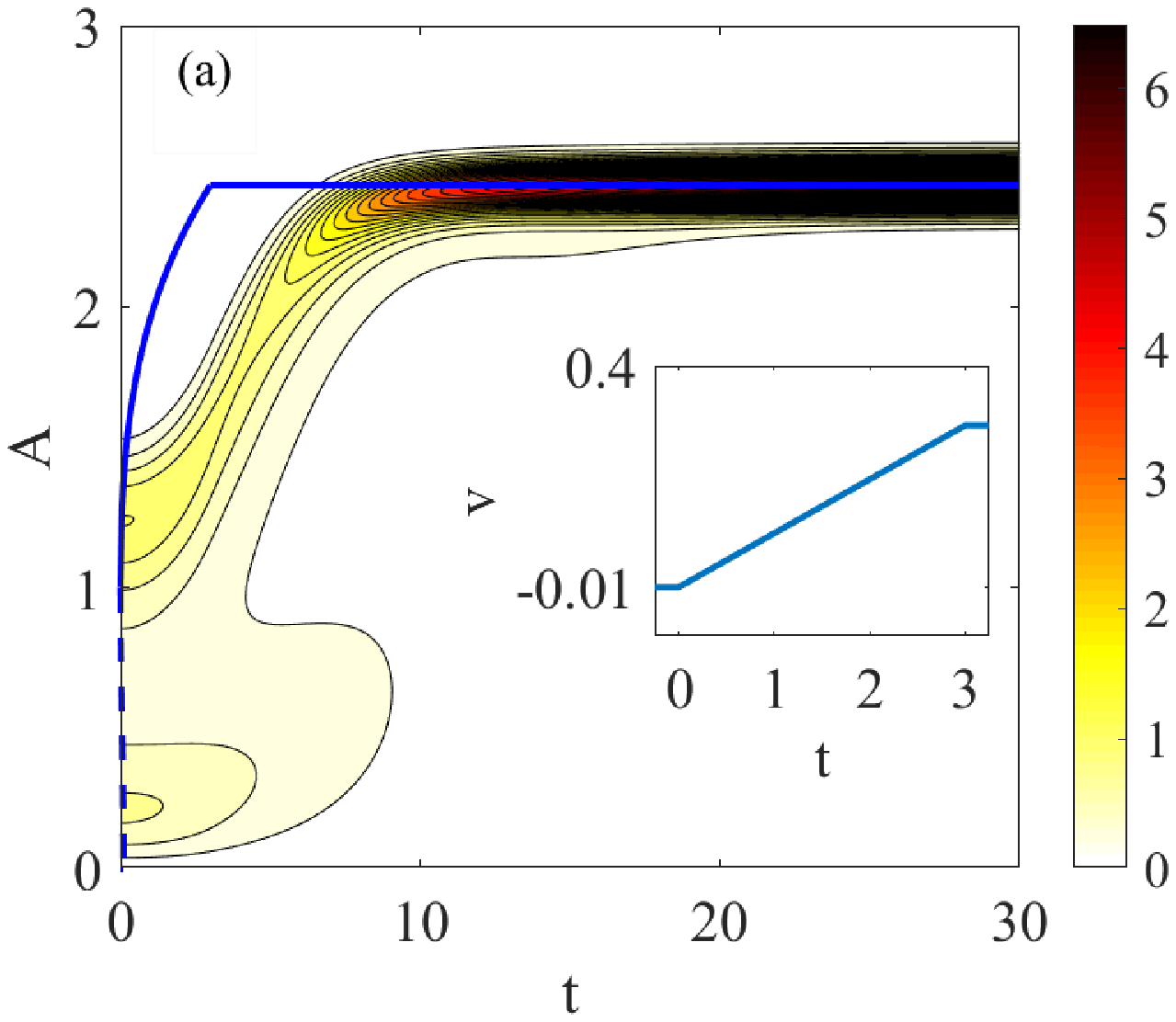}
	\includegraphics[width=8cm]{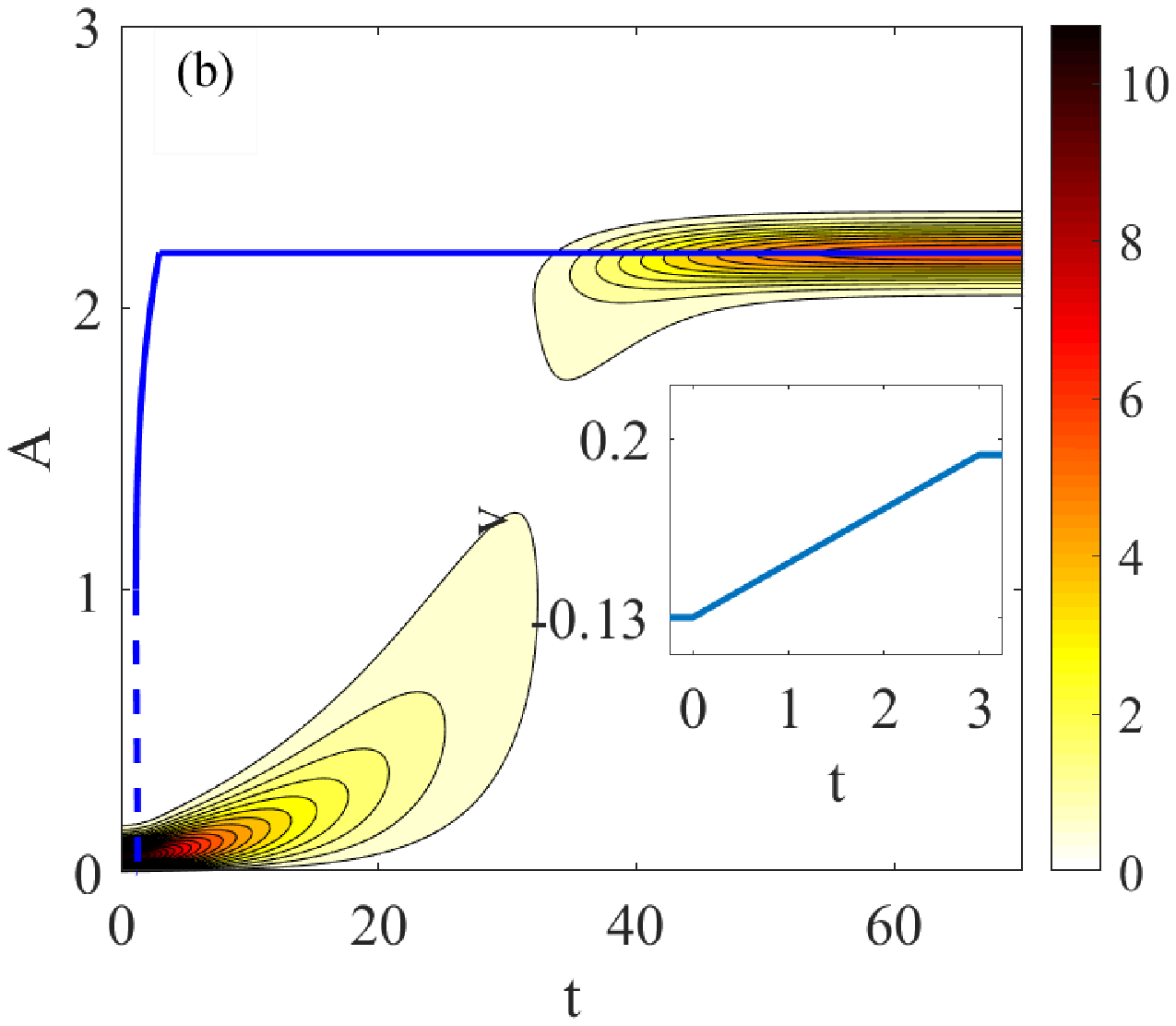}
	\includegraphics[width=8cm]{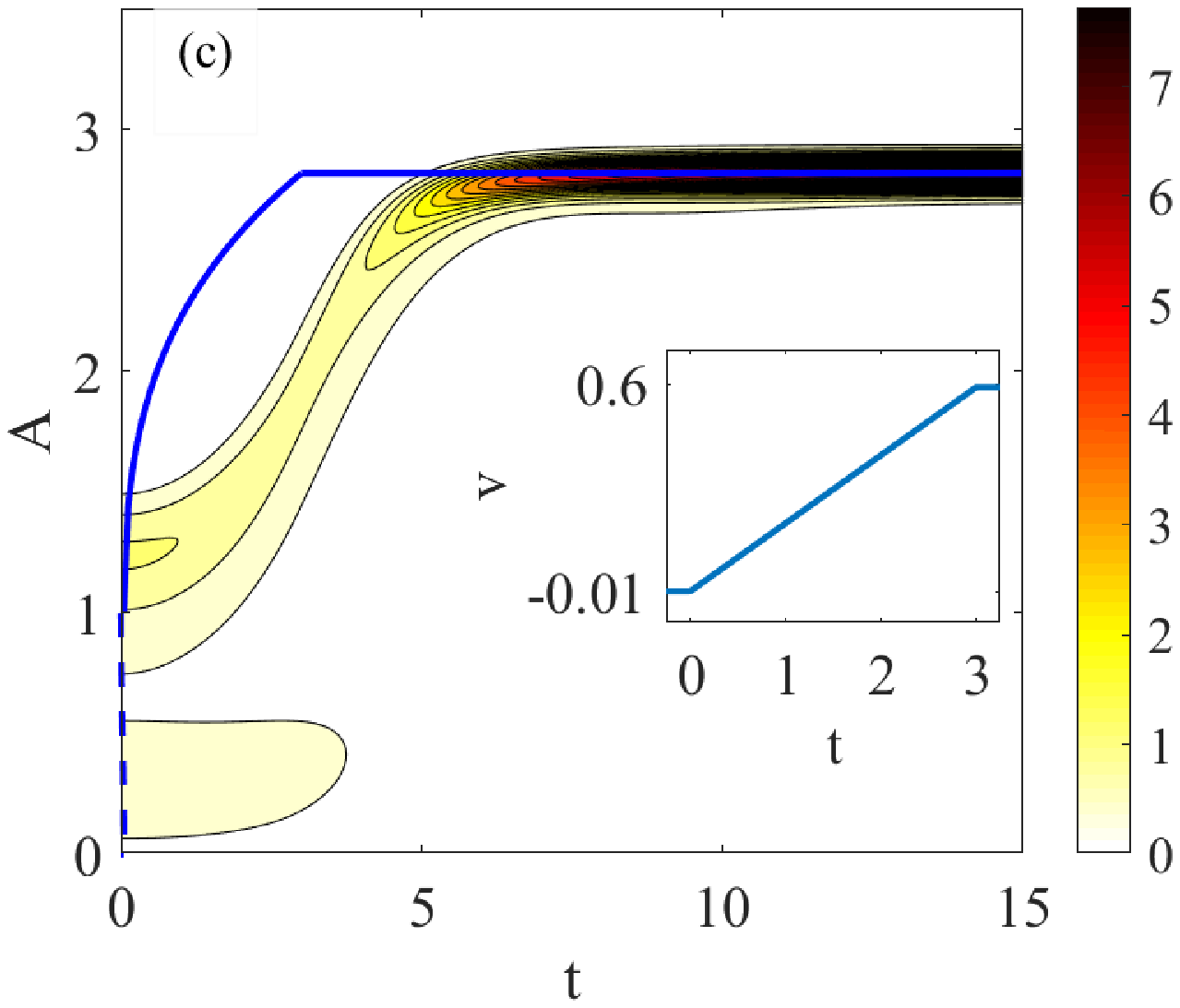}
	\includegraphics[width=8cm]{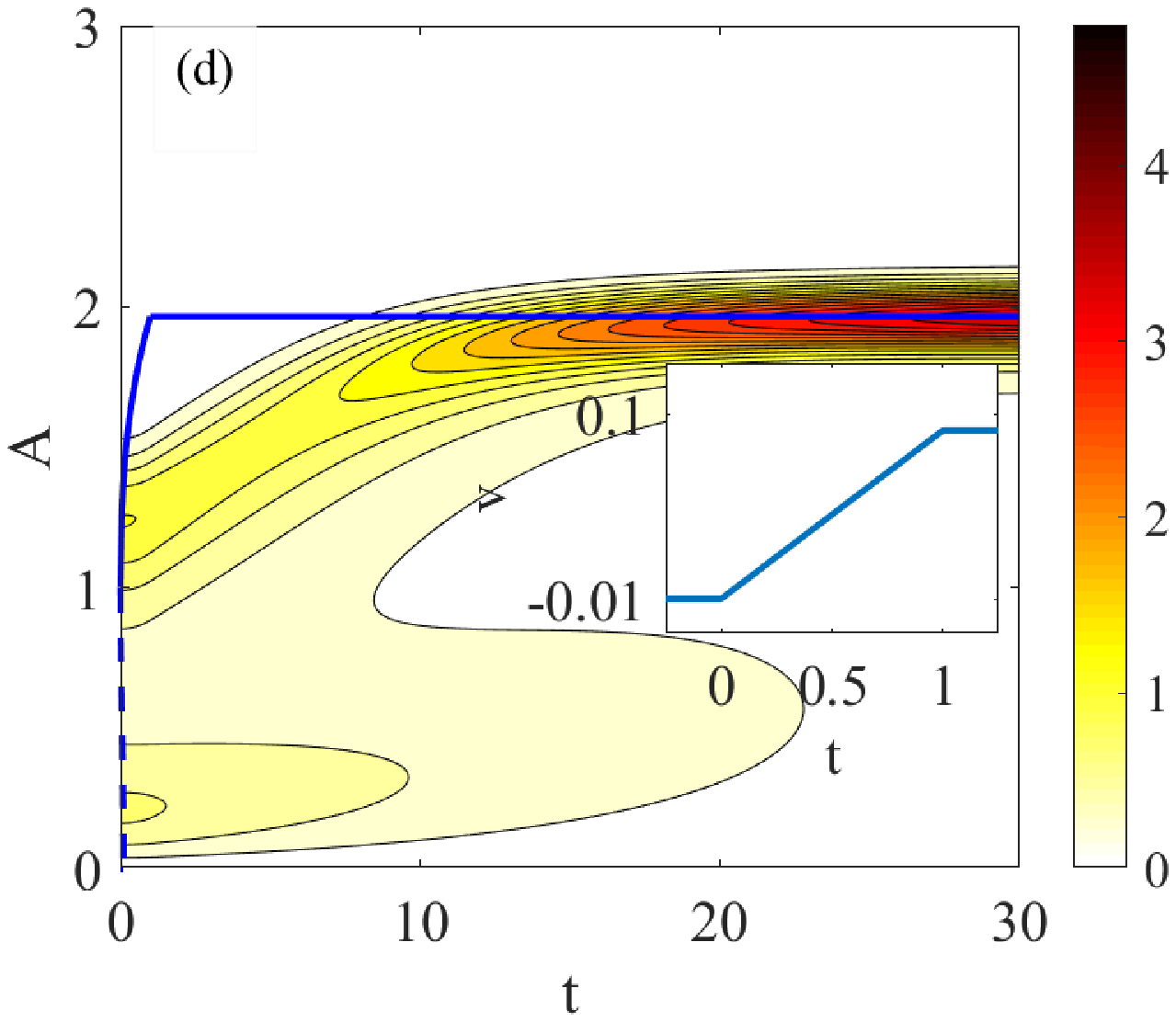}
	\caption{The effect of different initial value $v_0$, rate $R$ and changing time $t_c$ of the time-varying parameter $v(t)$ on the transient dynamical behavior under additive and multiplicative colored noise. Other parameters are $\beta_1=0.1$, $\beta_2=0.1$, $\omega_{0}=120\times 2\pi$, $D_1=6000$, $D_2=6000$, $\tau_1=0.007$, $\tau_2=0.001$. (a) $v_0=-0.01$, $R=0.1$, $t_c=3$; (b) $v_0=-0.13$, $R=0.1$, $t_c=3$; (c) $v_0=-0.01$, $R=0.2$, $t_c=3$; (d) $v_0=-0.01$, $R=0.1$, $t_c=1$.}
	\label{fig11}
\end{figure}

Using Figure \ref{fig11}(a) as a reference, Figures \ref{fig11}(b), \ref{fig11}(c) and \ref{fig11}(d) control the initial value, the ramp rate and the changing time of the variable respectively, to study the stochastic dynamical behavior of the system. The initial value $v_0$ has a great influence on the transient dynamical behavior of the system. When $v_0=-0.13$, the PDF passes through a very short bimodal state from the unimodal state to another and completes state switching; while $v_0=-0.01$, the PDF starts directly from the bimodal state and converts to the unimodal state after a relatively long time. The delay time of the rate-dependent tipping-delay phenomenon decreases with the increase of the rate $R$ or changing time $t_c$. This is consistent with the case where there is only a multiplicative colored noise excitation.

\section{Dependence of correlation time of noises on dynamical behaviors}
The white noise we are familiar with is a kind of noise with zero correlation time, but noise in real life often has a non-zero value. The correlation time, as a parameter of noise, may determine the dynamical behavior of the system. This section will discuss the impact of the correlation time of noises on the system response, especially for the tipping-delay phenomenon.

\begin{figure}[htbp]
	\centering
	\includegraphics[width=8cm]{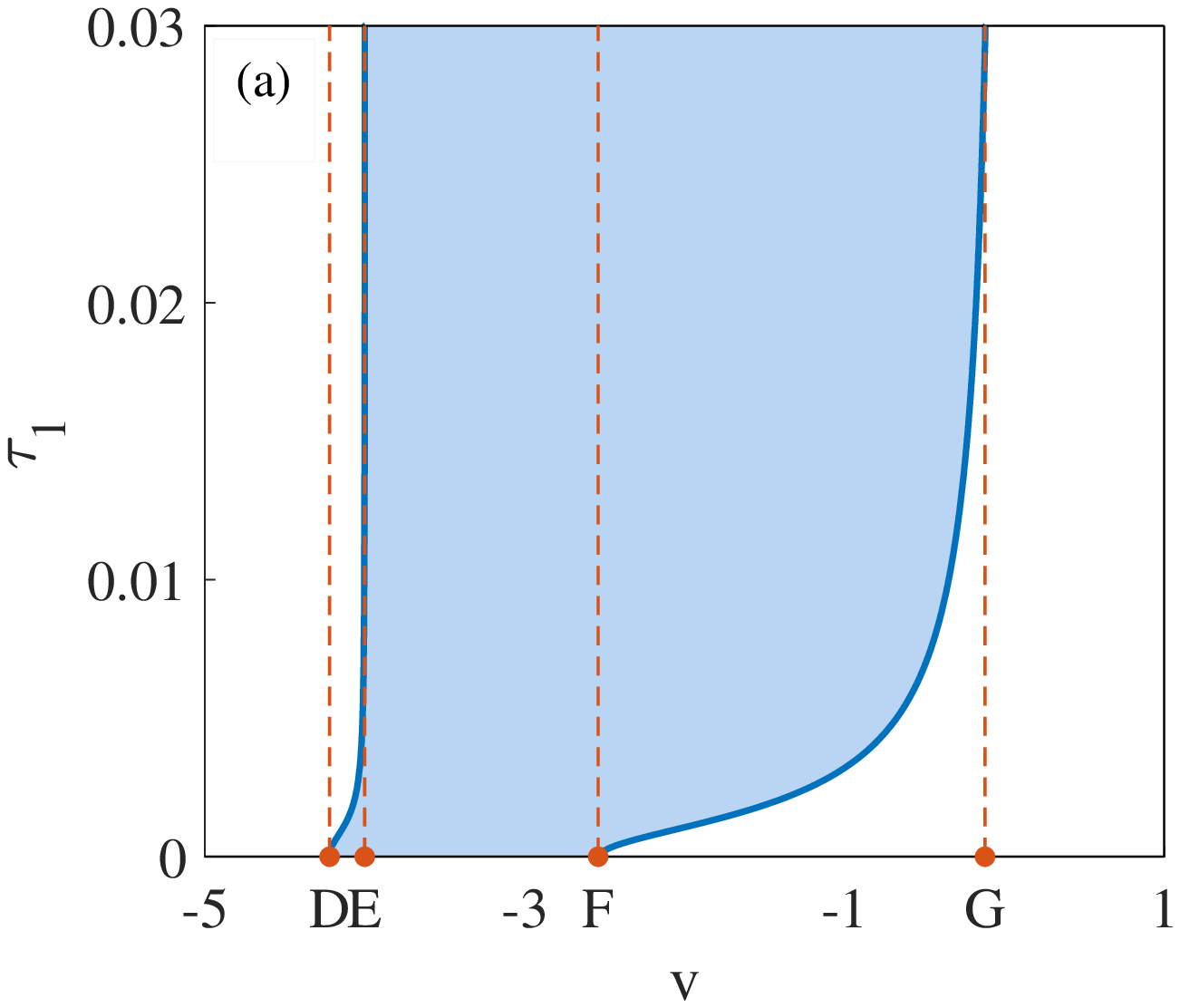}
	\includegraphics[width=8cm]{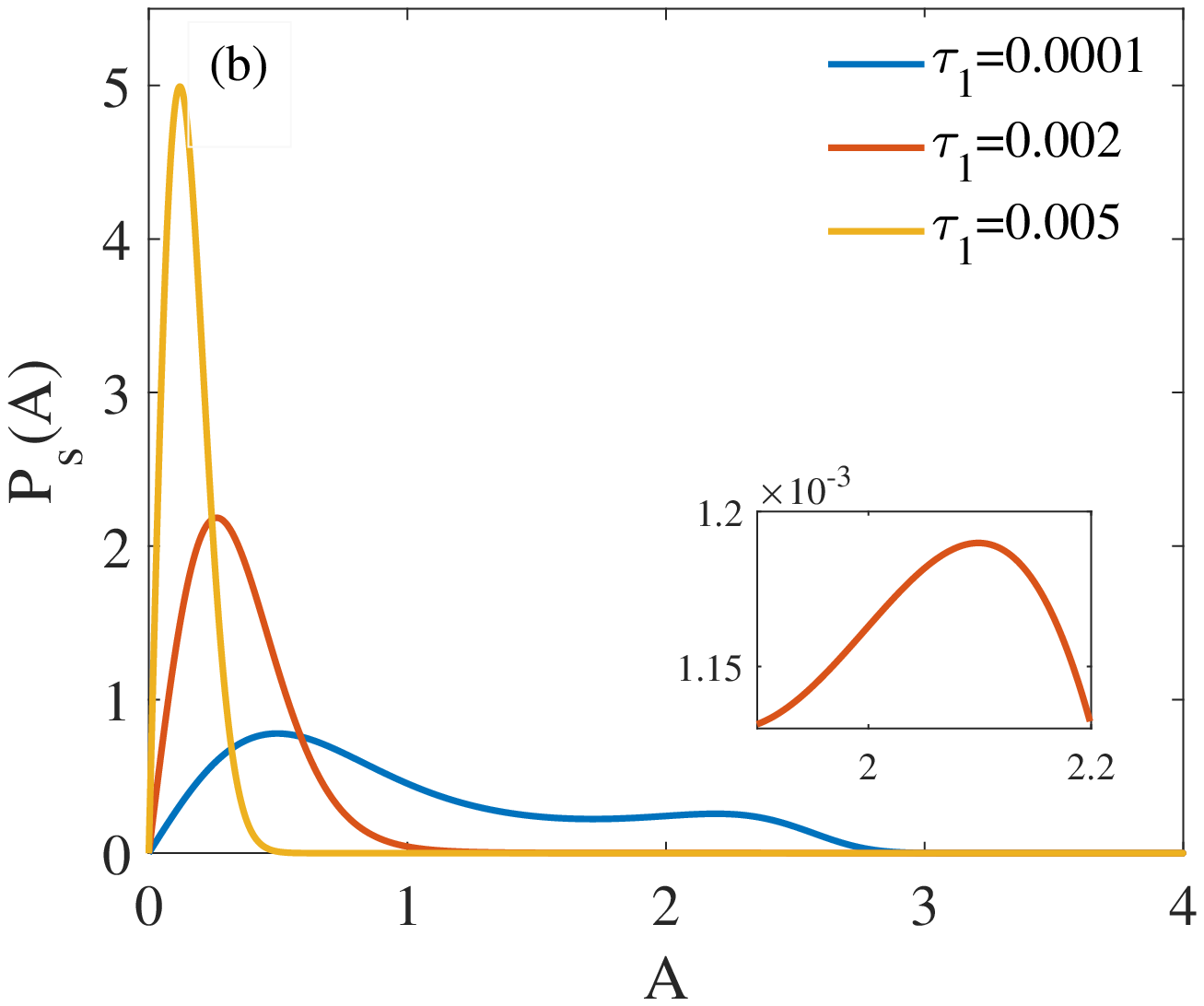}
	\caption{(a) Stochastic P-bifurcations on the parameter plane $(v,\tau_1)$ for systems disturbed only by additive colored noise for $\beta_1=8$, $\beta_2=2$, $\omega_0=120\times 2\pi$, $D_1=5\times10^5$; (b) The SPDF under different $\tau_1$ for systems disturbed only by additive colored noise at $v=-4.02\in(D,E)$.}
	\label{fig12}
\end{figure}

We first consider the system only disturbed by the additive colored noise. Figure \ref{fig12}(a) is a stochastic P-bifurcation diagram of the system on the parameter plane $(v,\tau_1)$. In Figure \ref{fig12}(a), the SPDF of the amplitude $A$ has a bimodal structure in the colored region and a unimodal one in the colorless region. When $v<D$, the value of $\tau_1$ does not affect the property of the PDF, and the PDF always maintains a unimodal state. Increasing the parameter $v$, PDF in the $v\in(D,E)$ interval with the increase of $\tau_1$, exhibits the single peak after the double peak. As $v$ increases, the bimodal region will expand in the direction of large $\tau_1$. Within $v\in(E,F)$, for any $\tau_1$, it exhibits a bimodal structure. We continue to increase the parameter $v$. When $v\in(F,G)$ the bimodal region becomes more and more narrow with increasing of $v$. At this time, as  $\tau_1$ increases, the PDF exhibits the first double peak after a single peak. For a further increase, when $v>G$, the bimodal region does not exist, and it always has a unimodal structure.

Figure \ref{fig12}(b) is SPDF for different correlation times $\tau_1$ at $v=-4.02\in(D,E)$. The SPDF changes from a bimodal structure to a unimodal one as $\tau_1$ increases. However, it is worth noting that when the correlation time is slightly larger (such as $\tau_1=0.002$), the SPDF has a bimodal structure, but the peak on the right is particularly small (see the internal enlarged view), and the particles are mainly concentrated at the left peak.

\begin{figure}[htbp]
	\centering
	\includegraphics[width=5cm]{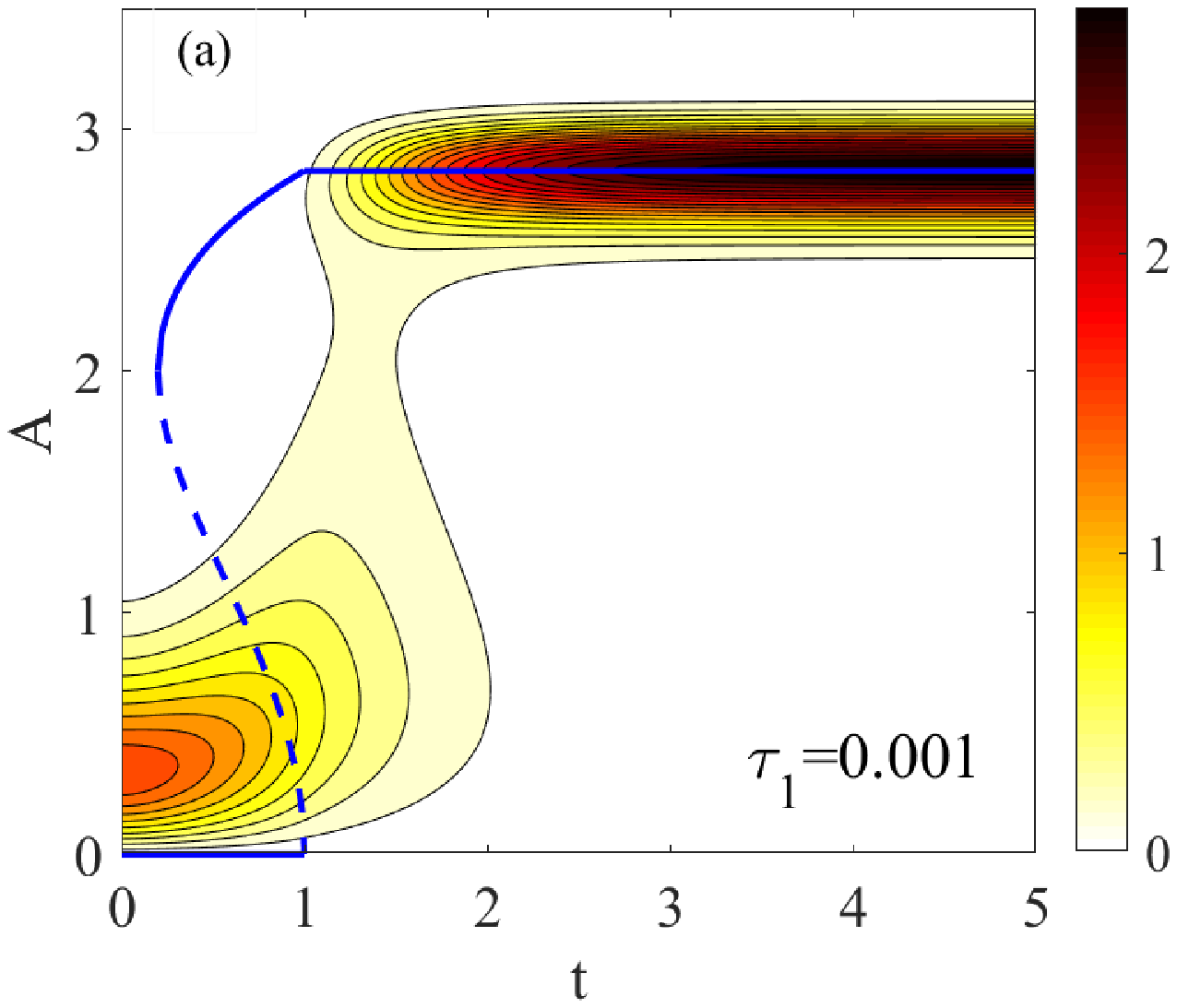}
	\includegraphics[width=5cm]{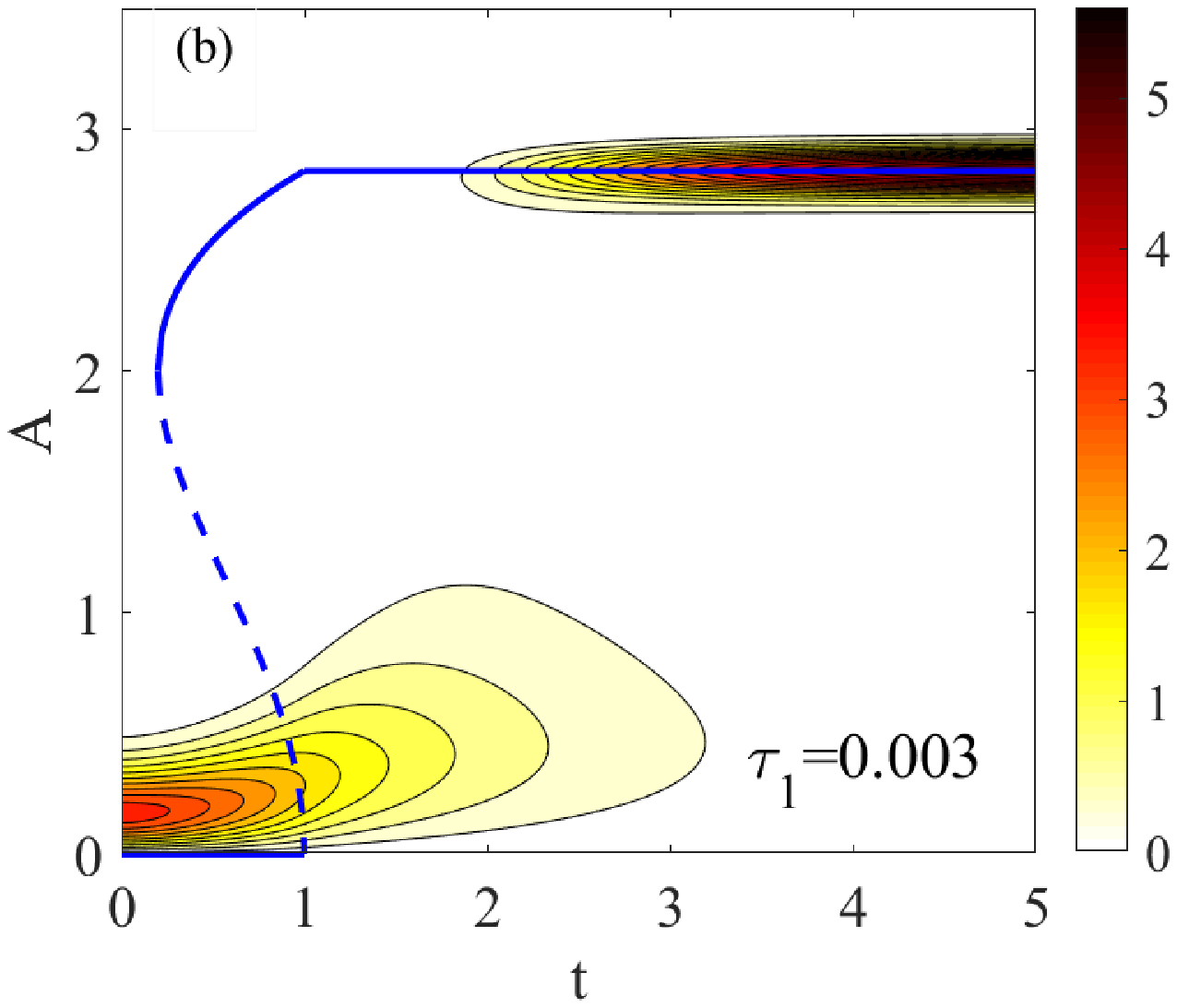}
	\includegraphics[width=5cm]{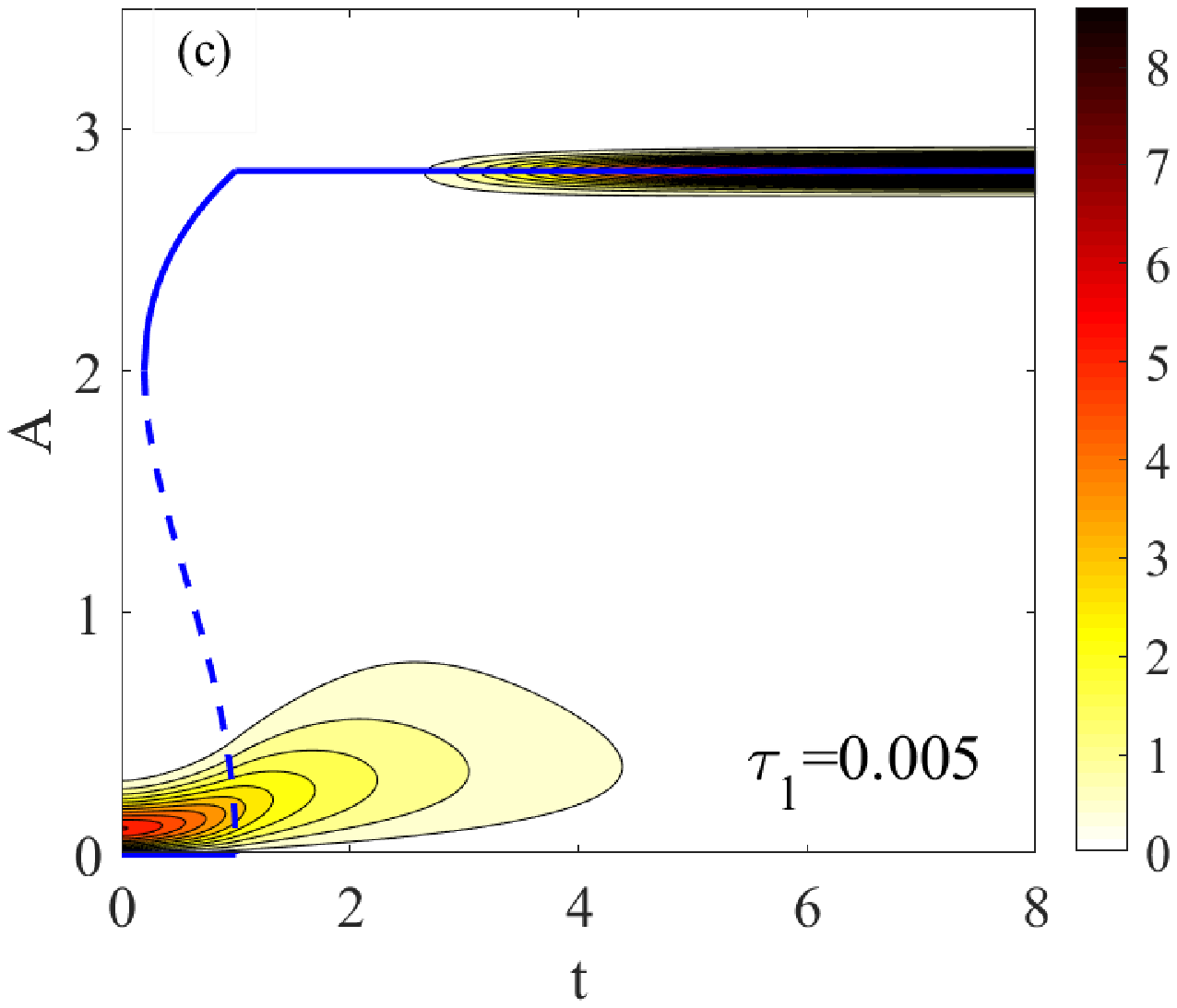}
	\includegraphics[width=5cm]{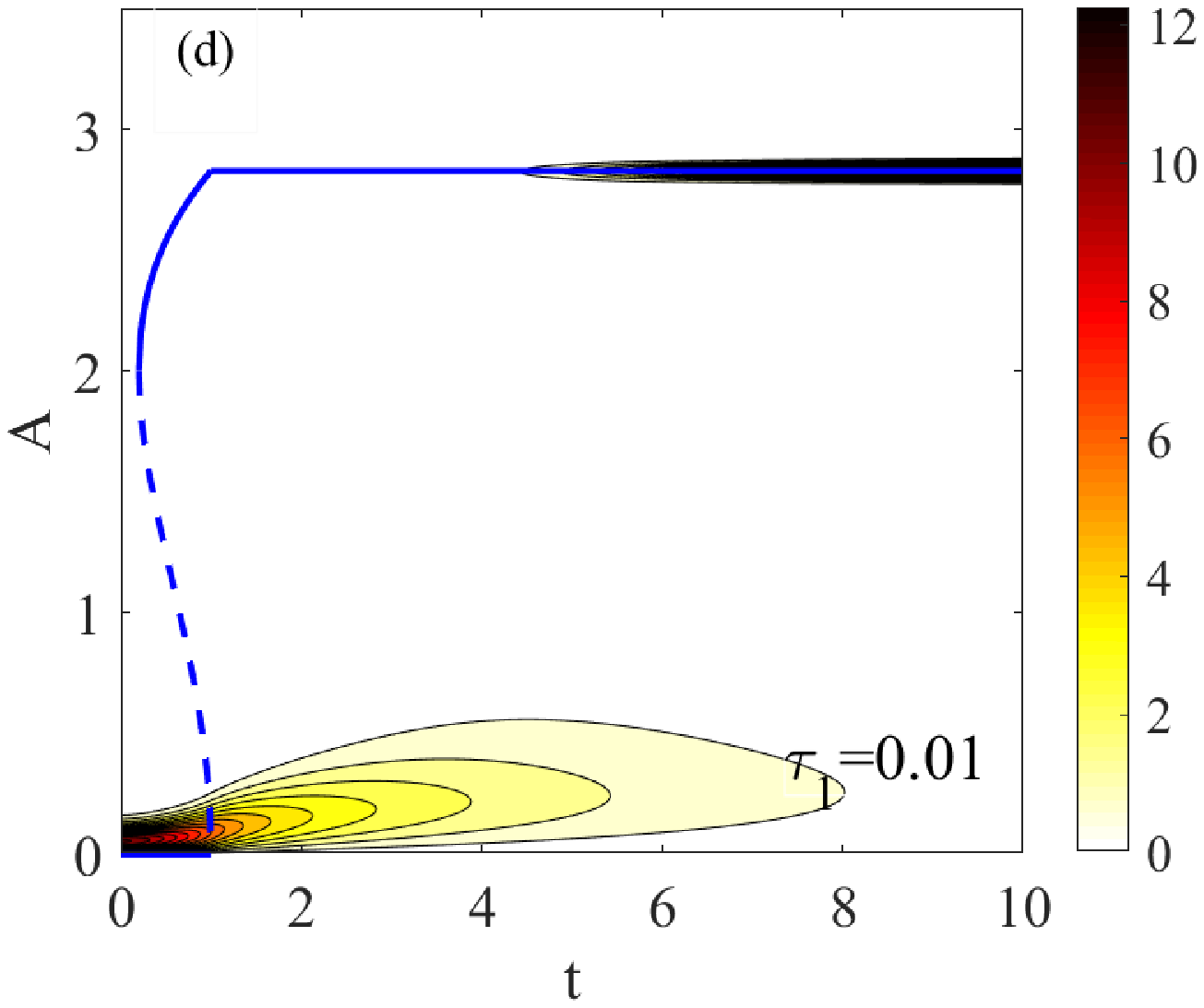}
	\includegraphics[width=5cm]{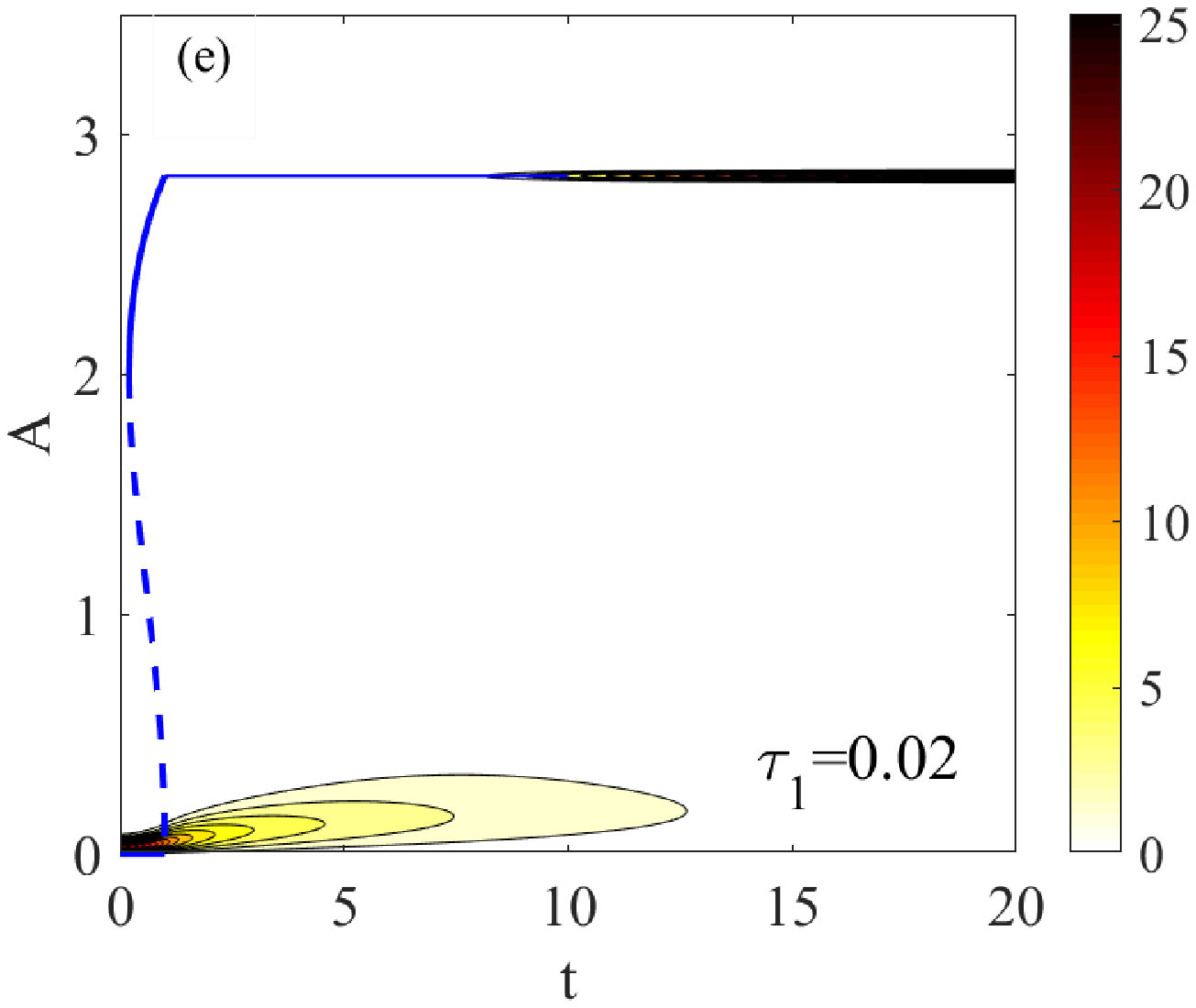}
	\includegraphics[width=5cm]{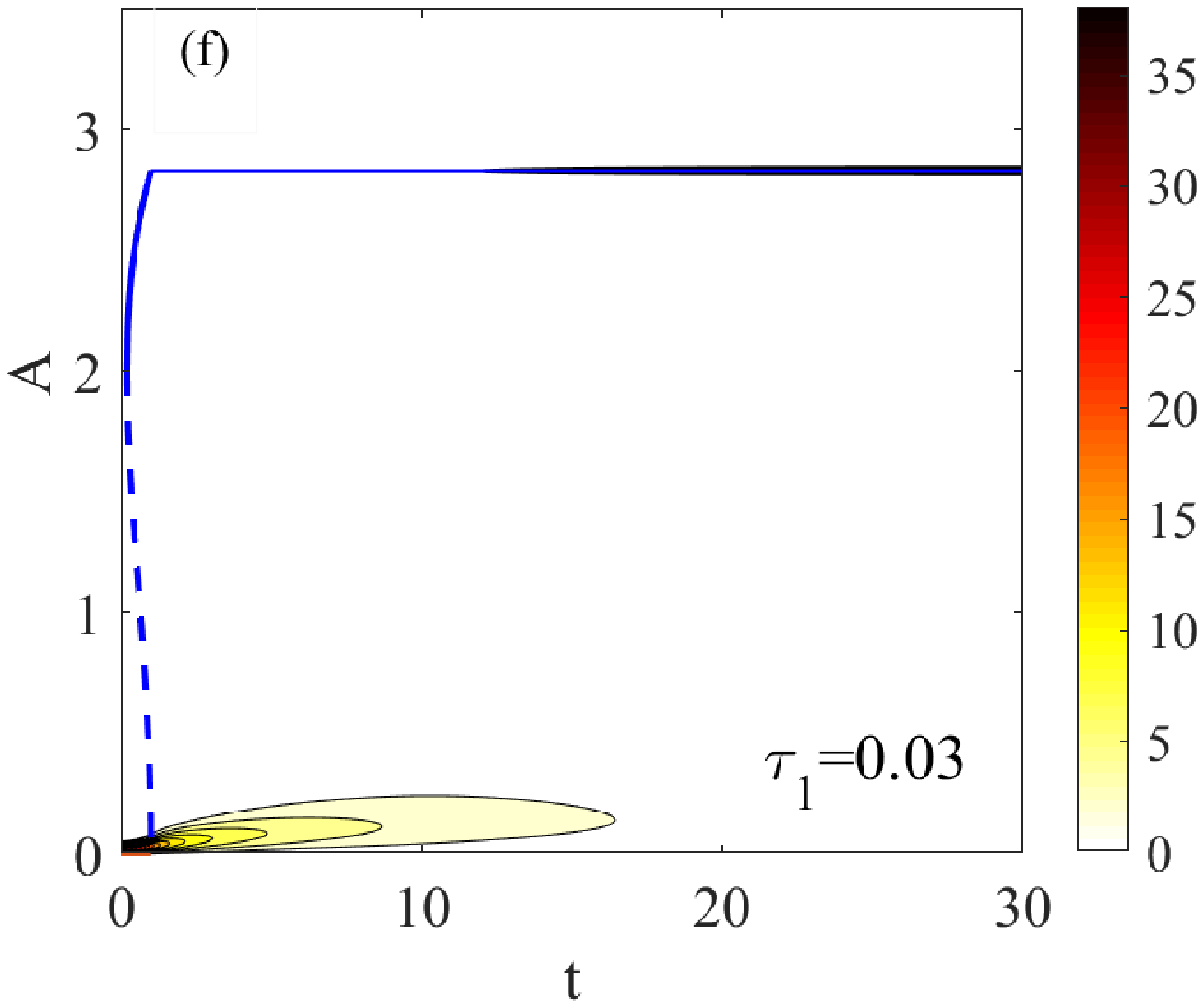}
	\caption{The effect of different $\tau_1$ on transient dynamical behavior under additive colored noise for $v_0=-5$, $R=5$, $t_c=1$. Other parameters are $\beta_1=8$, $\beta_2=2$, $\omega_0=120\times 2\pi$, $D_1=5\times10^5$.}
	\label{fig14}
\end{figure}

\begin{figure}[htbp]
	\centering
	\includegraphics[width=8cm]{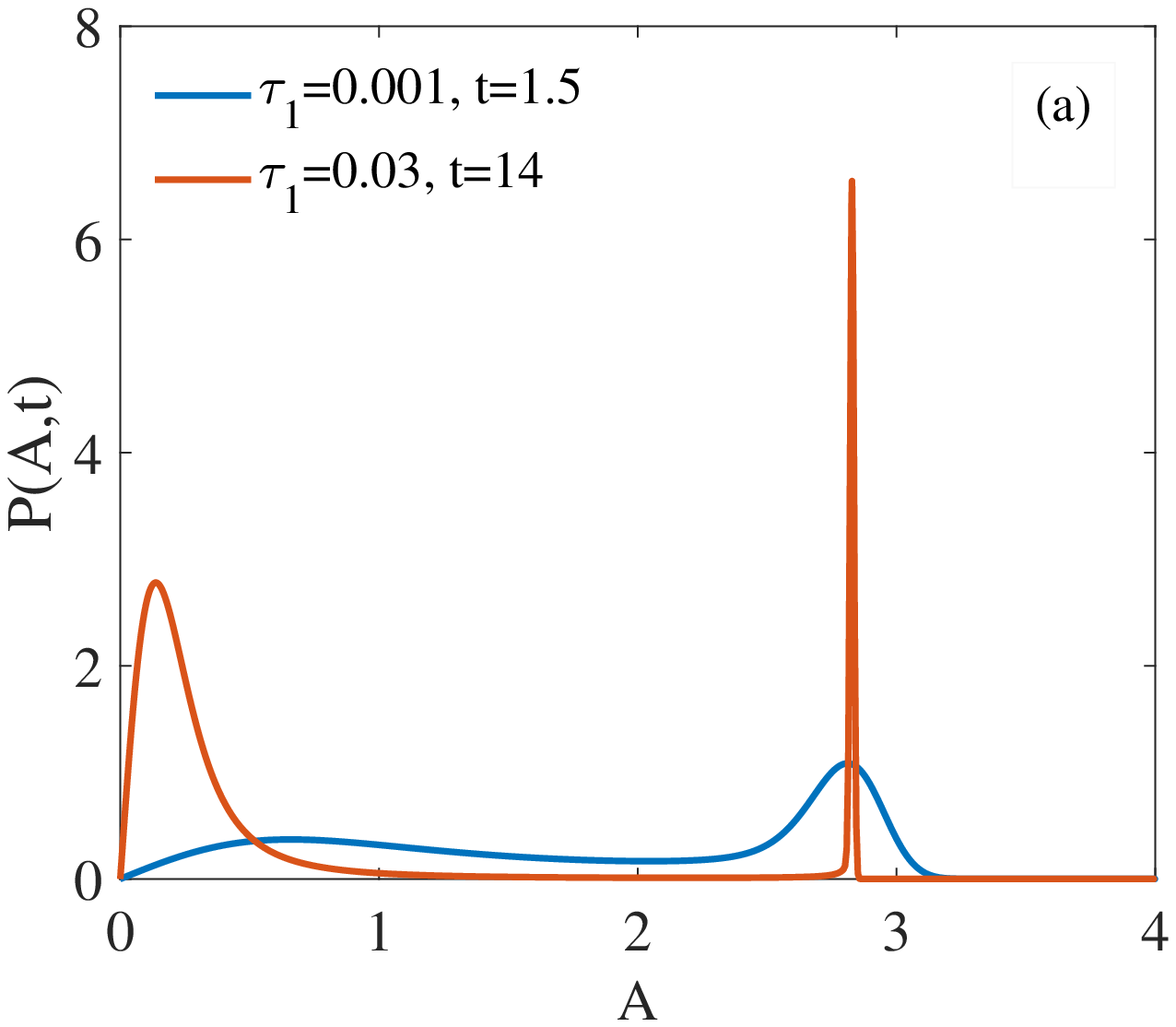}
	\includegraphics[width=8cm]{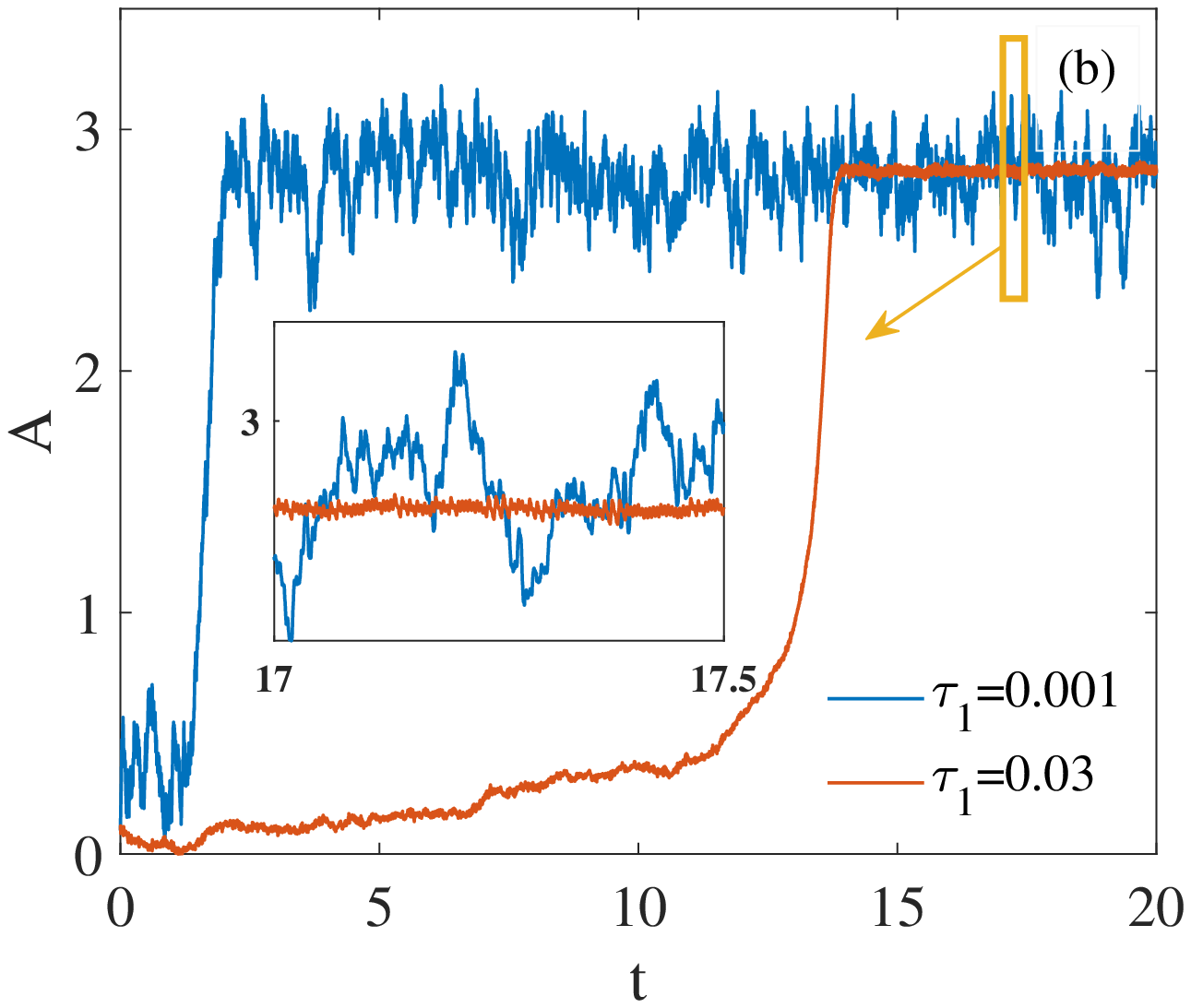}
	\caption{Diagrams under different $\tau_1$ for systems disturbed only by additive colored noise. Other parameters are $\beta_1=8$, $\beta_2=2$, $\omega_0=120\times 2\pi$, $D_1=5\times10^5$. (a) The PDF during the bimodal state; (b) The sample paths.}
	\label{fig15}
\end{figure}

Now we study the rate-dependent tipping-delay phenomenon of systems that are only excited by additive colored noise when $\tau_1$ takes different values. As can be seen from Figure \ref{fig14}, as $\tau_1$ increases, the line between the two stable regions gradually disappears, and the shape of each stable region is gradually compressed and flattened, and even concentrated in a high amplitude state to a line. This shows that the greater the correlation time of the noise, the smaller the probability that particles will be between two stable states when tipping occurs, and the more thoroughly the tipping transfer. This phenomenon is more obvious in Figure \ref{fig15}(a), which shows a PDF of the bimodal state during system tipping under different $\tau_1$. Moreover, when the correlation time is large, the fluctuation range of the particles near each stable point is small, which can be directly obtained from the sample path diagram of Figure \ref{fig15}(b). The blue line in  Figure \ref{fig14} is the result of the quasi-steady assumption. As a reference, it can be concluded that the larger the $\tau_1$, the larger the tipping-delay time. In addition, the duration of the bimodal state of the PDF becomes longer as $\tau_1$ increases.

\begin{figure}[htbp]
	\centering
	\includegraphics[width=8cm]{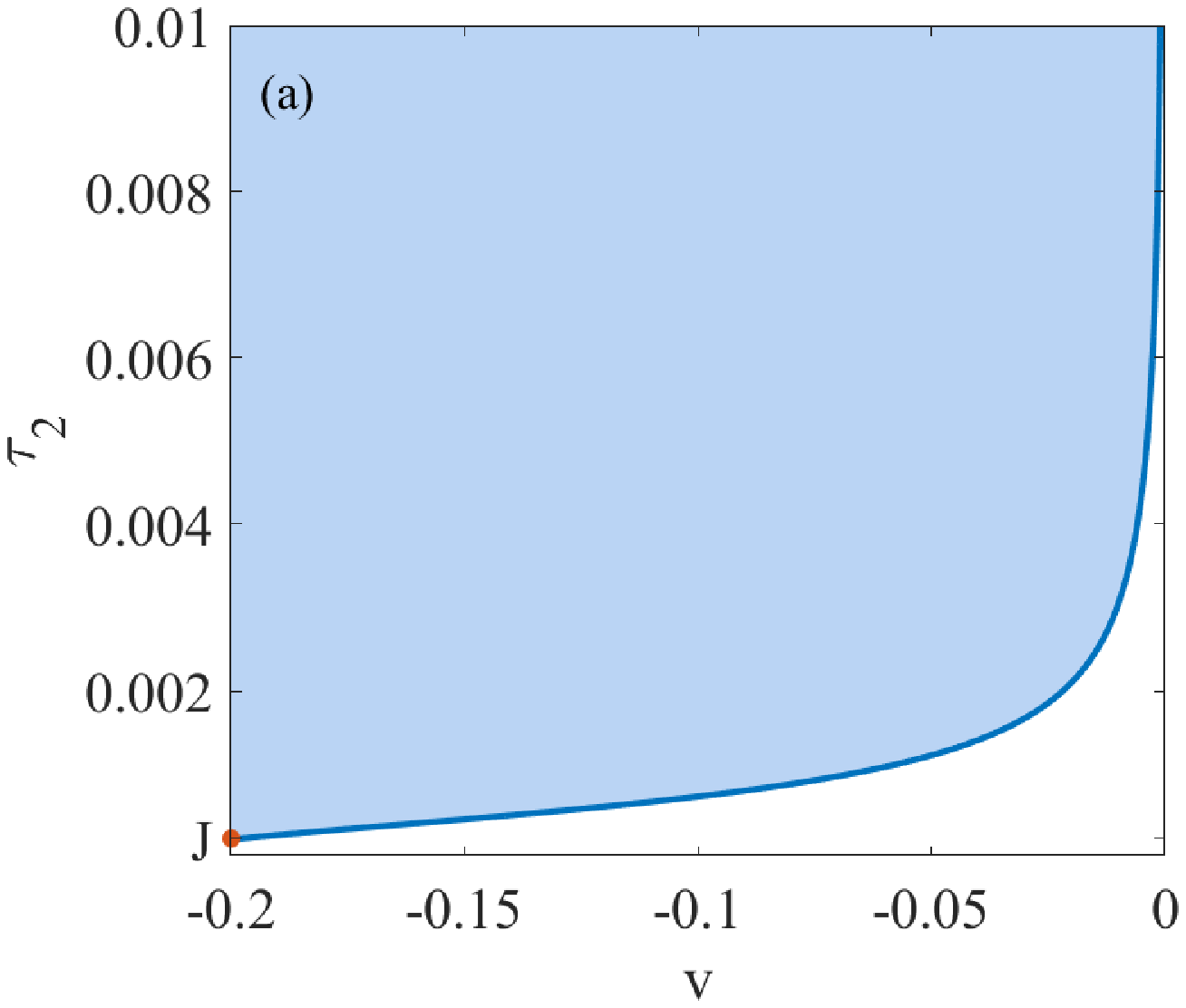}
	\includegraphics[width=8cm]{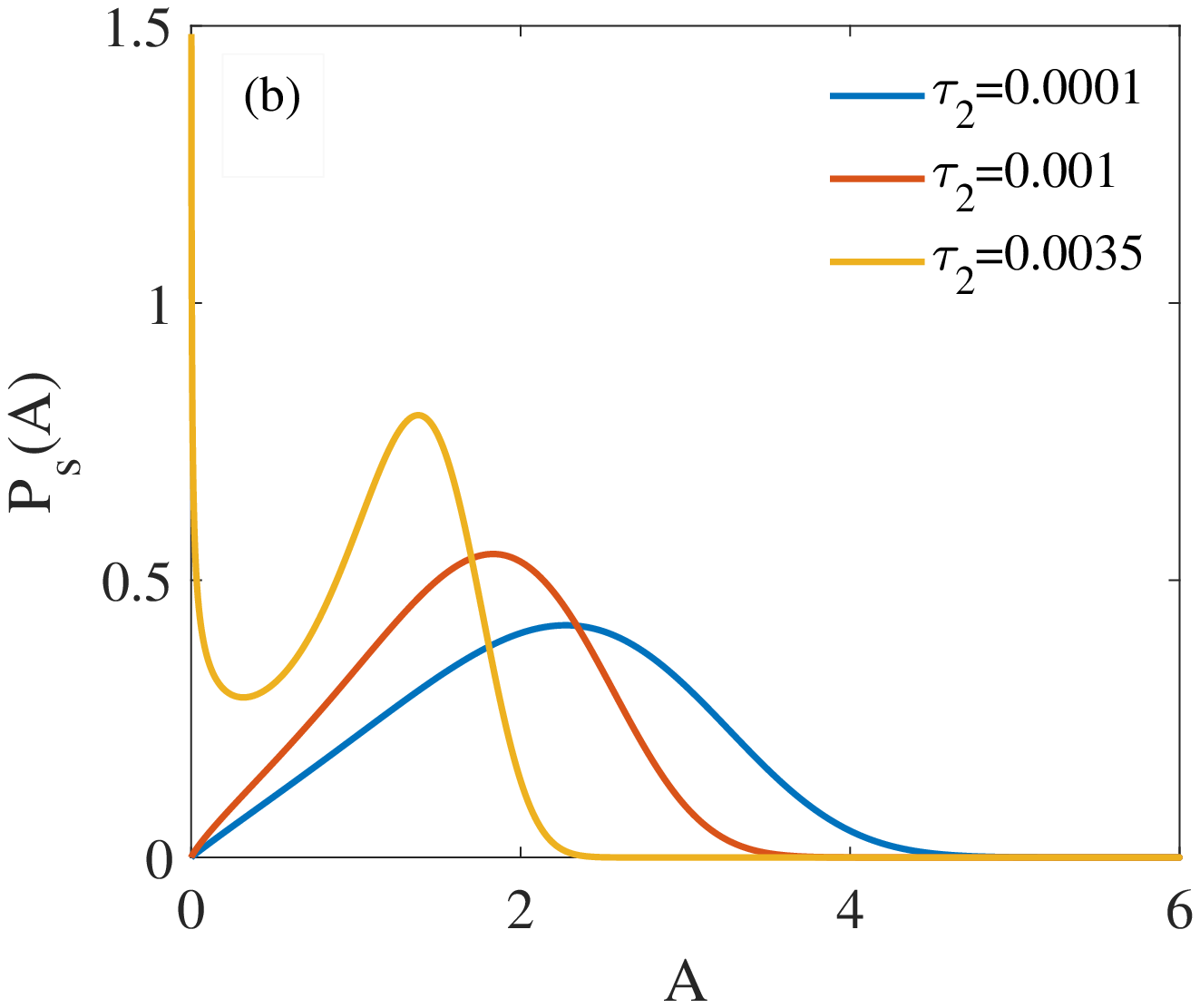}
	\caption{(a) Stochastic P-bifurcations on the parameter plane $(v,\tau_2)$ for systems disturbed only by multiplicative colored noise for $\beta_1=0.1$, $\beta_2=0.1$, $\omega_0=120\times2\pi$, $D_2=5\times10^5$; (b) The SPDF under different $\tau_2$ for systems disturbed only by multiplicative colored noise at $v=-0.01$.}
	\label{fig16}
\end{figure}

When only the multiplicative colored noise is active, the bifurcation diagram of $(v,\tau_2)$ is shown in Figure \ref{fig16}(a). At $v<0$, the SPDF of the amplitude has a structure of two extreme values in the colored region (as shown in Figure \ref{fig4}(b)) and a monopole structure in the colorless region. It is known from the formula that when $v>0$, the SPDF of the system is monostable for any correlation time. For very small $\tau_2$, the system always maintains a monostable structure regardless of the value of $v$. At $\tau_2>J$, as $v$ increases, the SPDF changes from a bipolar value to a unipolar value structure. When $\tau_2<J$, the SPDF always shows a unimodal state. Figure \ref{fig16}(b) is a SPDF at different correlation times $\tau_2$ at $v=-0.01$. At this time, the SPDF changes from a unipolar value structure to a bipolar one as $\tau_2$ increases.

\begin{figure}[htbp]
	\centering
	\includegraphics[width=8cm]{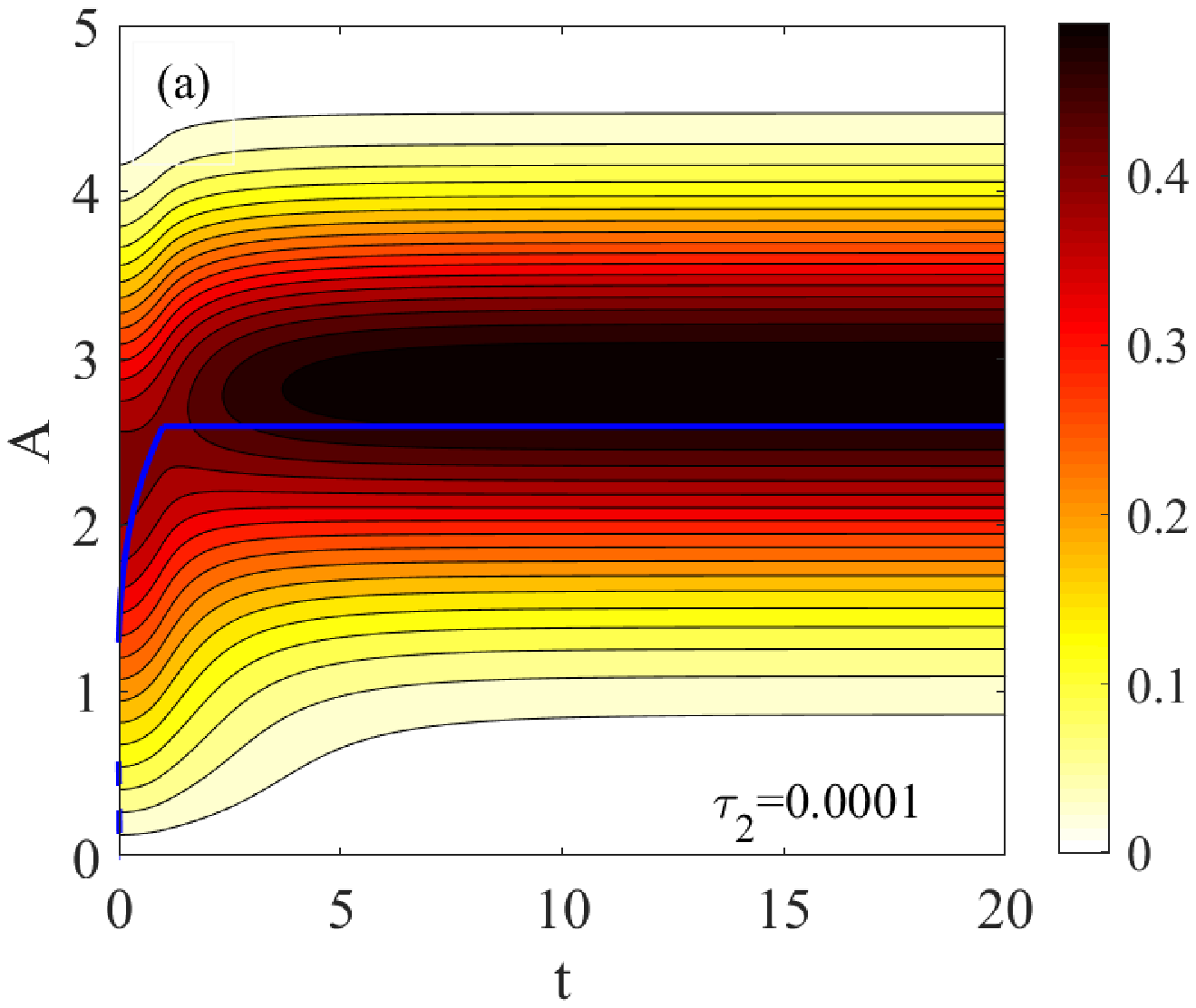}
	\includegraphics[width=8cm]{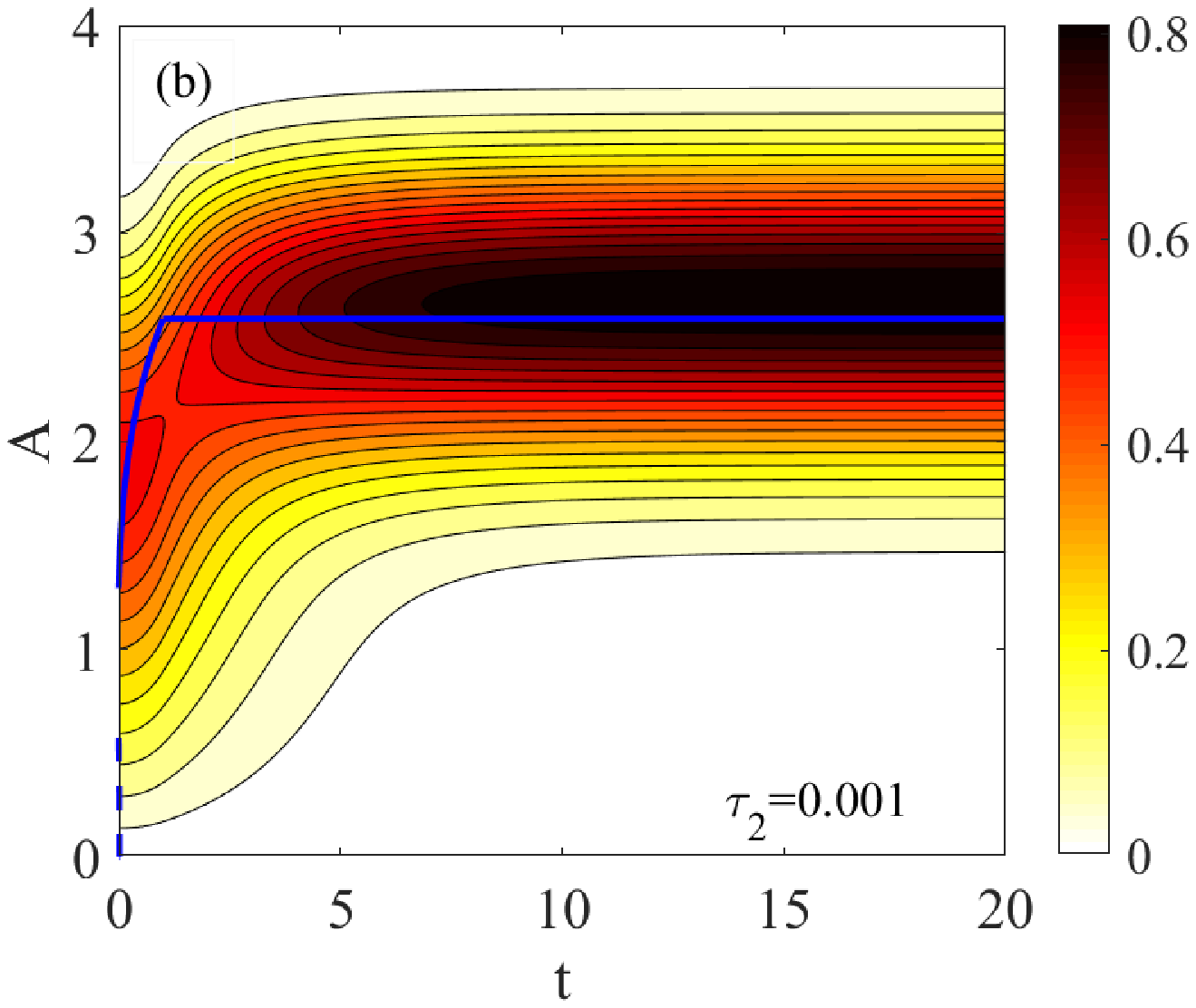}
	\includegraphics[width=8cm]{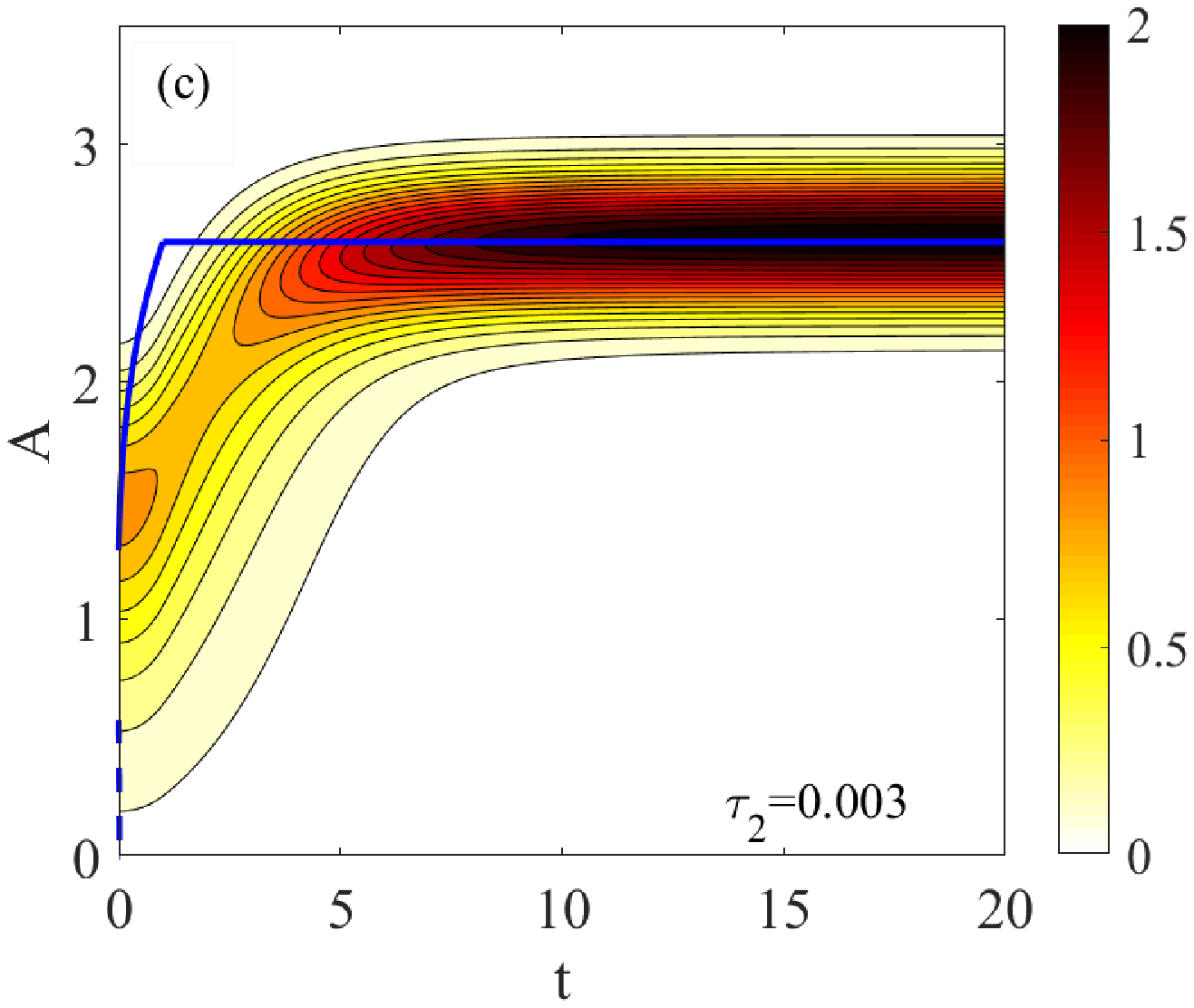}
	\includegraphics[width=8cm]{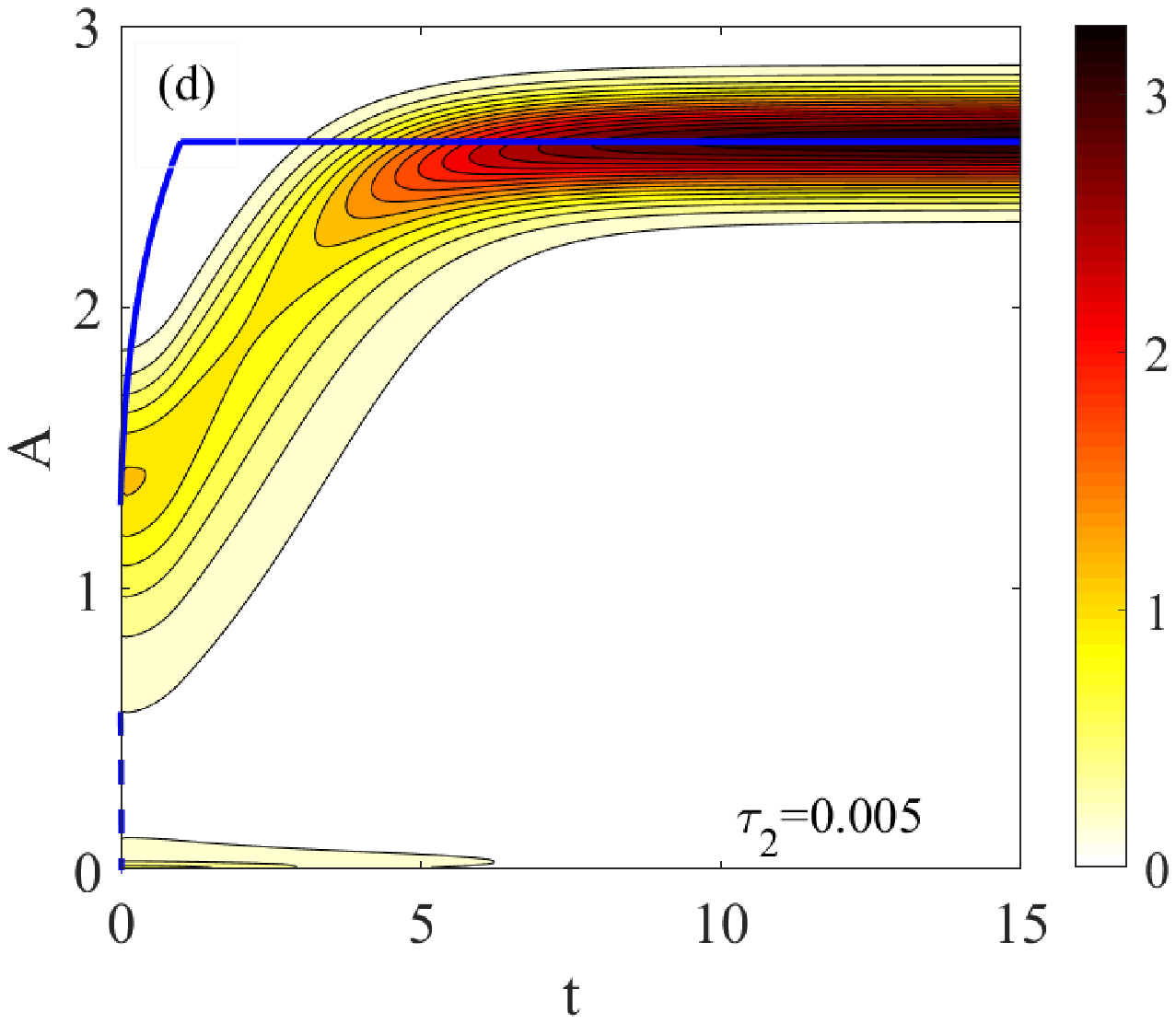}
	\caption{The effect of different $\tau_2$ on transient dynamical behavior under multiplicative colored noise for $v_0=-0.006$, $R=0.4$, $t_c=1$. Other parameters are $\beta_1=0.1$, $\beta_2=0.1$, $\omega_0=120\times2\pi$, $D_2=5\times10^5$.}
	\label{fig17}
\end{figure}

Figure \ref{fig17} shows the rate-dependent tipping-delay phenomenon of a system excited by multiplicative colored noise when $\tau_2$ takes different values. When $\tau_2$ is small, the PDF always maintains a unimodal state. At $\tau_2=0.005$, the PDF exhibits a bimodal to unimodal tipping phenomenon. As $\tau_2$ increases, the particle's fluctuation range near the stable region becomes smaller and the distribution becomes more concentrated. The blue lines are the quasi-steady results. As a reference, it can be concluded that the larger the $\tau_2$, the larger the tipping-delay time, which is consistent with the law of the additive case. It is worth noting that when $\tau_2$ is small, the maximum possible probability density of the finally arriving steady-state region is deviated, compared to the quasi-steady result due to the wide range of particle distribution.

When the system is excited by additive and multiplicative colored noise, the positive proportion relationship between the correlation time of noise and the tipping-delay time in the above case is still maintained, which is not discussed in detail here. In addition, the increase of the correlation time of noise makes the particles more concentrated near the stable point, which is also the main difference between colored noise and white noise.

\section{Conclusions}

This paper mainly studies a class of thermoacoustic systems with a subcritical Hopf bifurcation  structure. Considering a ramp-rate changing parameter and the colored noise excitations, the rate-dependent tipping-delay phenomenon is analyzed in detail. We employed the FPK equation obtained by the stochastic averaging method to study the transient dynamical behaviors of the thermoacoustic system. Then, the rate-dependent tipping-delay phenomenon of the thermoacoustic system is discussed. Through the above examination, several important results are uncovered. First, when we introduce the rate into the system, the tipping-delay phenomenon occurs. Second, the initial values, the ramp rate, the changing time, and the correlation time of noises have considerable influences on the rate-dependent tipping-delay phenomenon. Last, the result of the colored noise excitation is a generalization of the white noise case. Compared to the white noise, the relationship between the ramp rate and the delay time is still maintained in the case of colored noise. However, the main difference is that the particles tipping more thoroughly in terms of colored noise, which is further illustrated to study the dependence of the tipping-delay phenomenon on the correlation time of noises. Increasing the correlation time makes the particles more concentrated in the stable sets.

These results of our study can be employed to adhere the control of the dynamics of thermoacoustic systems. In the process of reducing the amplitude, it is not desirable to delay the tipping for too long in the thermoacoustic systems. Because long-term high amplitude vibration will cause irreversible damage to the system. Based on our results, we can choose an appropriate initial value, ramp rate, changing time or correlation time to shorten the tipping-delay time. In another way, if there is an early warning of the thermoacoustic instability, we can also consider a control of the parameters to extend the delay time. Then we use enough delay time period to avoid thermoacoustic instability completely. All in all, our work can provide a guiding significance for several practical problems in thermoacoustics.

\section*{Acknowledgments}
This work was supported by the National Natural Science Foundation of China (11772255), the Fundamental Research Funds for the Central Universities and the Research Funds for Interdisciplinary subject, NWPU.

\bibliography{tipping_ref}
\bibliographystyle{alpha}

\end{document}